\documentclass[preprint,journal]{vgtc}       % preprint (journal style)

\ifpdf%                                % if we use pdflatex
  \pdfoutput=1\relax                   % create PDFs from pdfLaTeX
  \usepackage{graphicx}                % allow us to embed graphics files
  \DeclareGraphicsExtensions{.pdf,.png,.jpg,.jpeg} % for pdflatex we expect .pdf, .png, or .jpg files
\else%                                 % else we use pure latex
  \usepackage{graphicx}                % allow us to embed graphics files
  \DeclareGraphicsExtensions{.eps}     % for pure latex we expect eps files
\fi%

\graphicspath{{figures/}{pictures/}{images/}{./}} % where to search for the images

\usepackage{microtype}                 % use micro-typography (slightly more compact, better to read)
\PassOptionsToPackage{warn}{textcomp}  % to address font issues with \textrightarrow
\usepackage{textcomp}                  % use better special symbols
\usepackage{mathptmx}                  % use matching math font
\usepackage{times}                     % we use Times as the main font
\usepackage{cite}                      % needed to automatically sort the references
\usepackage{tabu}                      % only used for the table example
\usepackage{booktabs}                  % only used for the table example
\usepackage{color}

\usepackage{soul}

\newcommand{\Mspace}{M}
\newcommand{\Rspace}{R}
\vgtccategory{Research}
\vgtcpapertype{application/design study}
 
\title{Towards replacing physical testing of granular materials \\ with a Topology-based Model}

\author{Aniketh Venkat, Attila Gyulassy, Graham Kosiba, Amitesh Maiti, Henry Reinstein, \\ Richard Gee, Peer-Timo Bremer, and Valerio Pascucci}
\authorfooter{
\item
 Aniketh Venkat, Attila Gyulassy, and Valerio Pascucci are with the SCI Institute, University of Utah. Email: anikethvenkat@sci.utah.edu.
\item
 Graham Kosiba, Amitesh Maiti, Henry Reinstein, Richard Gee, and Peer-Timo Bremer are with Lawrence Livermore National Laboratory. E-mail: bremer5@llnl.gov.
}

\shortauthortitle{Topological Pore Network Model}
\abstract{In the study of packed granular materials, the performance of a sample (e.g., the detonation of a high-energy explosive) often correlates to measurements of a fluid flowing through it. The “effective surface area,” the surface area accessible to the airflow, is typically measured using a permeametry apparatus that relates the flow conductance to the permeable surface area via the Carman-Kozeny equation. This equation allows calculating the flow rate of a fluid flowing through the granules packed in the sample for a given pressure drop. However, Carman-Kozeny makes inherent assumptions about tunnel shapes and flow paths that may not accurately hold in situations where the particles possess a wide distribution in shapes, sizes, and aspect ratios, as is true with many powdered systems of technological and commercial interest. To address this challenge, we replicate these measurements virtually on micro-CT images of the powdered material, introducing a new Pore Network Model based on the skeleton of the Morse-Smale complex. Pores are identified as basins of the complex, their incidence encodes adjacency, and the conductivity of the capillary between them is computed from the cross-section at their interface. We build and solve a resistive network to compute an approximate laminar fluid flow through the pore structure. We provide two means of estimating flow-permeable surface area: (i) by direct computation of conductivity, and (ii) by identifying dead-ends in the flow coupled with isosurface extraction and the application of the Carman-Kozeny equation, with the aim of establishing consistency over a range of particle shapes, sizes, porosity levels, and void distribution patterns.%
} % end of abstract


\CCScatlist{ % not used in journal version
 \CCScat{K.6.1}{Management of Computing and Information Systems}%
{Project and People Management}{Life Cycle};
 \CCScat{K.7.m}{The Computing Profession}{Miscellaneous}{Ethics}
}

\teaser{
  \centering
  \includegraphics[width=\linewidth]{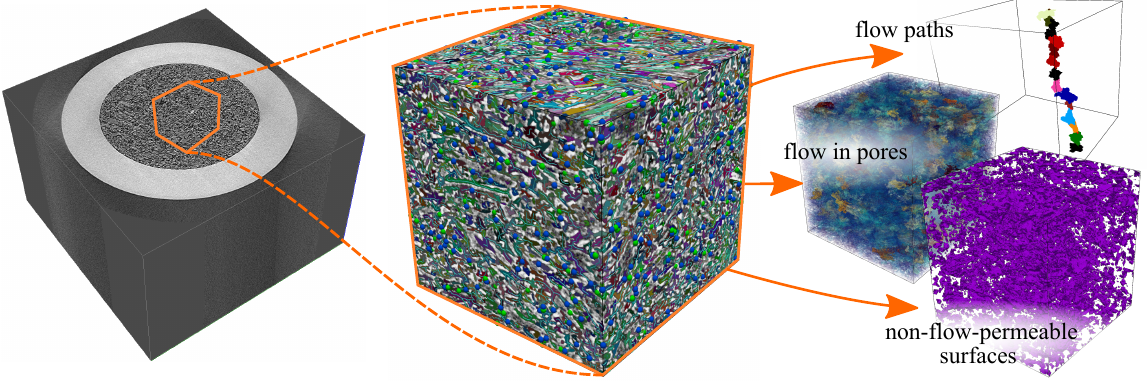}
  \caption{We derive a pore network model from topological decomposition and connectivity to estimate the flow properties through a packed powder bed imaged by micro-CT. From left to right: Regions are selected from the raw micro-CT images, and the pore network model is computed; The fluid flow is solved by converting it to a resistive network, which allows the analysis of flow paths, flow through pores, and flow-permeable surface area, all of which correlate to the performance characteristics of porous solids.}
  \label{fig:teaser}
}

\renewcommand{\manuscriptnotetxt}{}

\vgtcinsertpkg
\usepackage{subcaption}

\begin{document}


\firstsection{Introduction}

\maketitle

Diverse industries such as pharmaceuticals, cosmetics, paints, textiles, minerals, packaging, structural and building materials (e.g., cement), and energetic materials (e.g., initiating powders) involve the need to quantify the flow-permeable surface area of porous compounds. Commercially available permeametry apparatuses, such as the traditional one from Fisher Scientific or a more modern version from Micromeritics carry out such measurement by flowing a fluid (air) through the powdered sample and then directly or indirectly relating the measured flow conductance (i.e., the ratio of flow rate to pressure-drop across the sample) to a permeable surface area of the powder through the Carman-Kozeny (CK) equation~\cite{Kozeny1927},~\cite{Carman1938}, \cite{Carman1956}. However, the CK equation at its core is based on a few inherent assumptions about the tunnel shapes and flow paths, which real systems could significantly deviate from, especially in cases where particles are far from spherical shapes, e.g., possessing needle-like morphology~\cite{ZEPEDARUIZ2006461}, as is known for PETN and other energetic initiators. Given that there is no obvious way to “extend” the Carman-Kozeny analysis in a general case of arbitrary shape and size distribution of particles packed randomly, it is essential to develop an alternative, direct approach to determining the flow-permeable surface area. 

This work attempts to achieve the above goal by imaging the porous sample material by micro-CT (i.e., Computed Tomography at micron or better resolution), followed by segmentation of the resulting micro-CT image into grains and voids, which can provide direct insight into the porous structure and the expected flow rate through the sample. Laminar flow of a viscous fluid through a porous material can be described by the Navier-Stokes equation, which is both computationally expensive, and difficult to extend to the imaged domains. As a result, there has been significant interest in simplified models, called Pore Network Models (PNM)~\cite{Blunt2001}, where the void space of a material is decomposed into distinct pores, connected by throats. By careful calculation of the conductance of each pore-throat-pore edge and by satisfying the conservation of mass of the Stokes equation, an approximate flow rate can be computed for the entire network.

Our new approach for analyzing porous materials extends this class of solutions by directly utilizing the topology of the distance function from the material/void interface. Using the Morse-Smale complex, our approach couples the volumetric decomposition of the pores with a 1-skeleton connecting them. 
%However, existing techniques have a number of short-comings by either processing scans one slide at a time which can severely overestimate the flow~\cite{Gunther} or making assumptions on the shape and size of the particles~\cite{PNM}.
%Here, we introduce a new approach based on analyzing the topology of the distance functions to the material interface. 
%Using a combination of image processing and topological segmentation techniques, we decompose the void space into a set of {\it reservoirs} connected by {\it capillaries}. 
We then compute the effective conductance of each connection using Poiseulle's law~\cite{Sutera1993}, measuring the interface between pores and the 1-skeleton paths. We interpret the resulting network as an electrical resistor network and use nodal analysis to find the flow and pressure drop between pores and the sample at large. 
%prethe effective volumetric flow rate of the entire sample, pressure drop between pores, and volum we can apply the  flow equation to estimate the conductance of each reservoir and the material as a whole. 
This not only enables us to identify dead-ends in the material but also presents two computational ways of estimating the flow rate, and hence the flow-permeable surface area, that would be measured in a Fisher Apparatus. We validate our results on simple examples, and compare our estimates to experimentally measured values for packed spheres with varying size distributions, and packed granular materials with varying grain sizes and shapes.

Our contributions in detail are:
\begin{itemize}
    \item A topological algorithm to compute the conductance of a pore-throat-pore connection, find the flow and pressure drop between pores
    \item A demonstration that despite irregular particle shapes and sizes, there exist no significant amount of non-flow-permeable area (due to dead-ends)
    \item Two independent virtual measurements of the flow rate through a granular material; and
    \item Validation of our technique on analytical examples, calibration data consisting of packed spheres, and High Explosives materials of interest to our collaborators.
\end{itemize}

\section{Background}
\begin{figure}[tb!]
 \centering % avoid the use of \begin{center}...\end{center} and use \centering instead (more compact)
 \includegraphics[width=.6\linewidth]{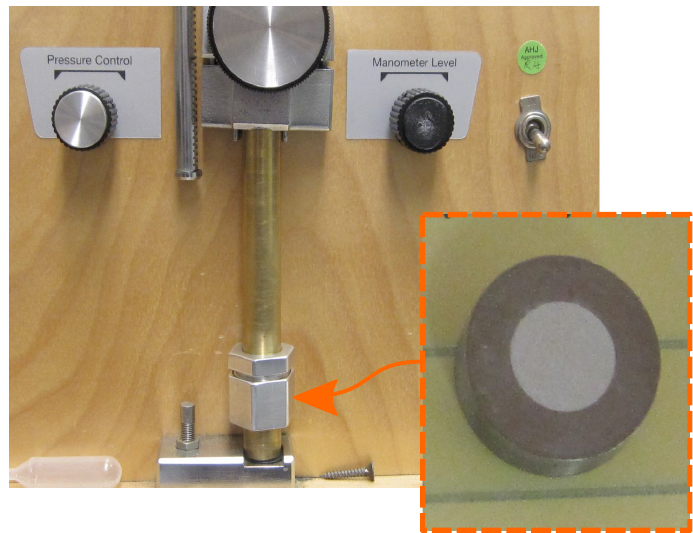}
 \caption{A granular material, here polydispersed spheres, is packed into a cylindrical tube (inset). The packed material is placed in the Fisher apparatus, which applies a pressure gradient, and measures the volume flow rate of air moving through the sample. }
 \label{fig:fisher}
\end{figure}

\begin{figure}[tb!]
 \centering % avoid the use of \begin{center}...\end{center} and use \centering instead (more compact)
 \includegraphics[width=.8\linewidth]{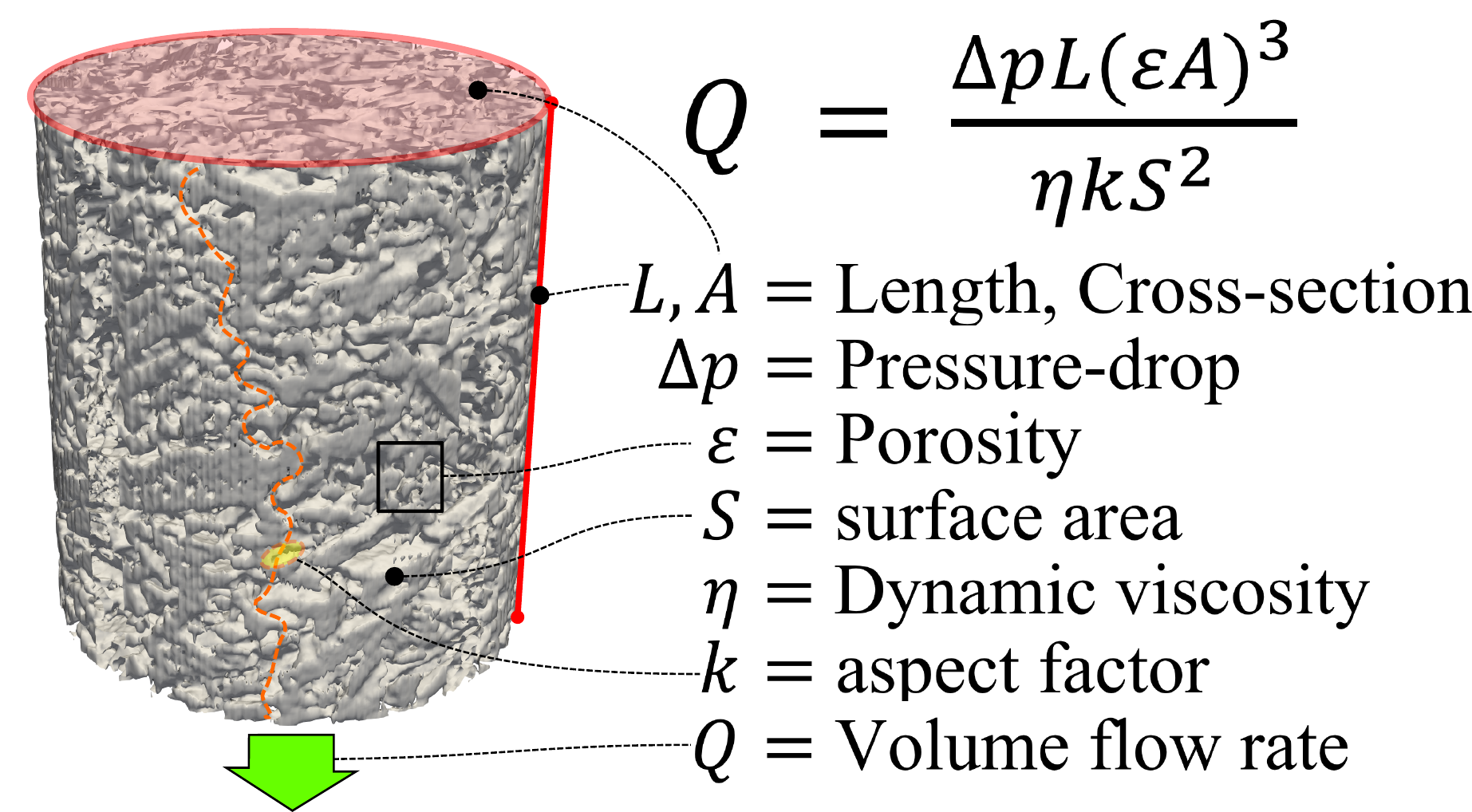}
 \caption{The variables involved in the Carman-Kozeny equation are illustrated for a packed porous material sample in a cylindrical container. The sides of the cylinder are enclosed in a container, with the induced pressure difference and flow only in the vertical axis of the cylinder. }
 \label{fig:carmankoseny}
\end{figure}

%\paragraph{Properties of porous materials} 
%A porous material has solid matter, typically formed by grains, and void space, the space between the solid particles. Given a container of length $L$ and cross-sectional area $A$, its volume $V$ is given by $V = LA$. The porosity $\epsilon$ of a material measures the fraction of $V$ that is composed of voids. A pressure gradient $\Delta p$ is applied across opposing ends, such that air acts as a viscous fluid flowing through the void structure, with viscosity $\eta$. The Carman-Kozeny equation (\autoref{fig:carmankoseny}) relates the volumetric flow rate across the sample to the geometry of the sample container, the change in pressure, the porosity of the material, the dynamic viscosity of air, and the flow-permeable surface area $S$.
A porous material is composed of solid matter, typically formed of grains, and voids, the space between the solid particles. The Fisher apparatus, ~\autoref{fig:fisher}, measures fluid flow properties across a prepared sample, relating the flow to material attributes. If a porous powder sample is packed into a cylindrical container of length $L$ and cross-sectional area $A$, its volume is given by $V = LA$. The porosity $\epsilon$ measures the fraction of $V$ that is composed of voids. When a pressure-drop $\Delta p$ is applied across opposing ends of the cylindrical sample, air flows as a viscous fluid through the void structure. In one form, the Carman-Kozeny equation (\autoref{eqn:carmankoseny} or \autoref{fig:carmankoseny}) relates the volumetric flow rate ($Q$) across the sample to the geometry of the sample container, the pressure-drop, the porosity of the material, the dynamic viscosity of air, and the flow-permeable surface area $S$, as follows:
\begin{equation}\label{eqn:carmankoseny}
Q= \frac{\Delta p L(\epsilon A)^3}{\eta k S^2 } 
\end{equation}

\autoref{eqn:carmankoseny} could be solved for the specific surface area ($SSA$), i.e., $SSA = S/M$ (where $M$ is the sample mass), and one obtains the result:
\begin{equation}\label{eqn:SSA}
SSA = \frac{S}{M} = \sqrt{\frac{\epsilon^3 A \Delta p}{(1 - \epsilon)^2 L k \eta Q \rho_S^2 } }
\end{equation}

where $\rho_S$ is the density of the solid. One often expresses the above result in terms of an effective spherical diameter ($d$):

\begin{equation}\label{eqn:spherical_diameter}
d = \frac{6}{\rho_S SSA} = 6\sqrt{\frac{(1 - \epsilon)^2 L k \eta Q}{\epsilon^3 A \Delta p}} 
\end{equation}

%Specifically, the equation relates the volumetric flow rate inversely proportional to the square of the flow-permeable surface area; therefore all else being held constant, decreasing the particle size, increases the flow-permeable surface area, decreases the expected measured flow rate across the sample. While this equation has been empirically verified for a number of materials, the extent of its applicability is not yet known, for instance, to what extent it can predict the relationship for materials with uneven grain size distribution or shapes. The aspect factor $k$, has empirically been measured to be around $5$ for packed spheres, as a product $k = k_0*k_1$ of a \textit{shape factor} set to 2 for spherical particles and a \textit{tortuosity factor}, which is the square of the effective path length a fluid particle travels through the material divided by the length of the container, $L_e/L$.

According to \autoref{eqn:carmankoseny}, the volumetric flow rate $Q$ is inversely proportional to the square of the flow-permeable surface area. Therefore, at a given level of porosity, the smaller the particle size, the higher the flow-permeable surface area, which leads to a decrease in the flow rate across the sample. While \autoref{eqn:carmankoseny} has been empirically verified for several materials, the extent of its applicability is not yet known. For instance, through many experiments on different powder systems, the aspect factor $k$ has been empirically determined to be around 5 \cite{Allen1997}.  However, $k$ is a product of two factors, i.e., $k = k_0 k_1$, where $k_0$ is a shape factor, which depends on the shape of the cross-section of flow tunnels, and $k_1$ the tortuosity factor, which is an enhanced path-length ratio when fluid particles meander through connected voids, rather than travel through a straight line parallel to the cylinder’s axis. Given a wide range of possible shapes and size distributions of particles and flow tunnels, it would not be surprising to encounter systems for which $k$ is significantly different from 5. However, it is not possible to determine $k$ for such systems without an independent method of determining the specific surface area.

One area of great interest to us is a special class of energetic materials known as High Explosives (HEs). Certain HEs, used as initiators, are often used in powder form. Common examples include PETN and HMX. It has been well-established that the higher the flow-permeable $SSA$, the higher the efficiency of initiation. With age, such powders are known to coarsen (i.e., loss in $SSA$), which lowers initiation efficiency~\cite{Maiti2017}. This strong dependence of performance makes an accurate determination of flow-permeable $SSA$ an important problem in the HE community. 

\begin{figure*}[tb!]
 \centering % avoid the use of \begin{center}...\end{center} and use \centering instead (more compact)
 \includegraphics[width=\linewidth]{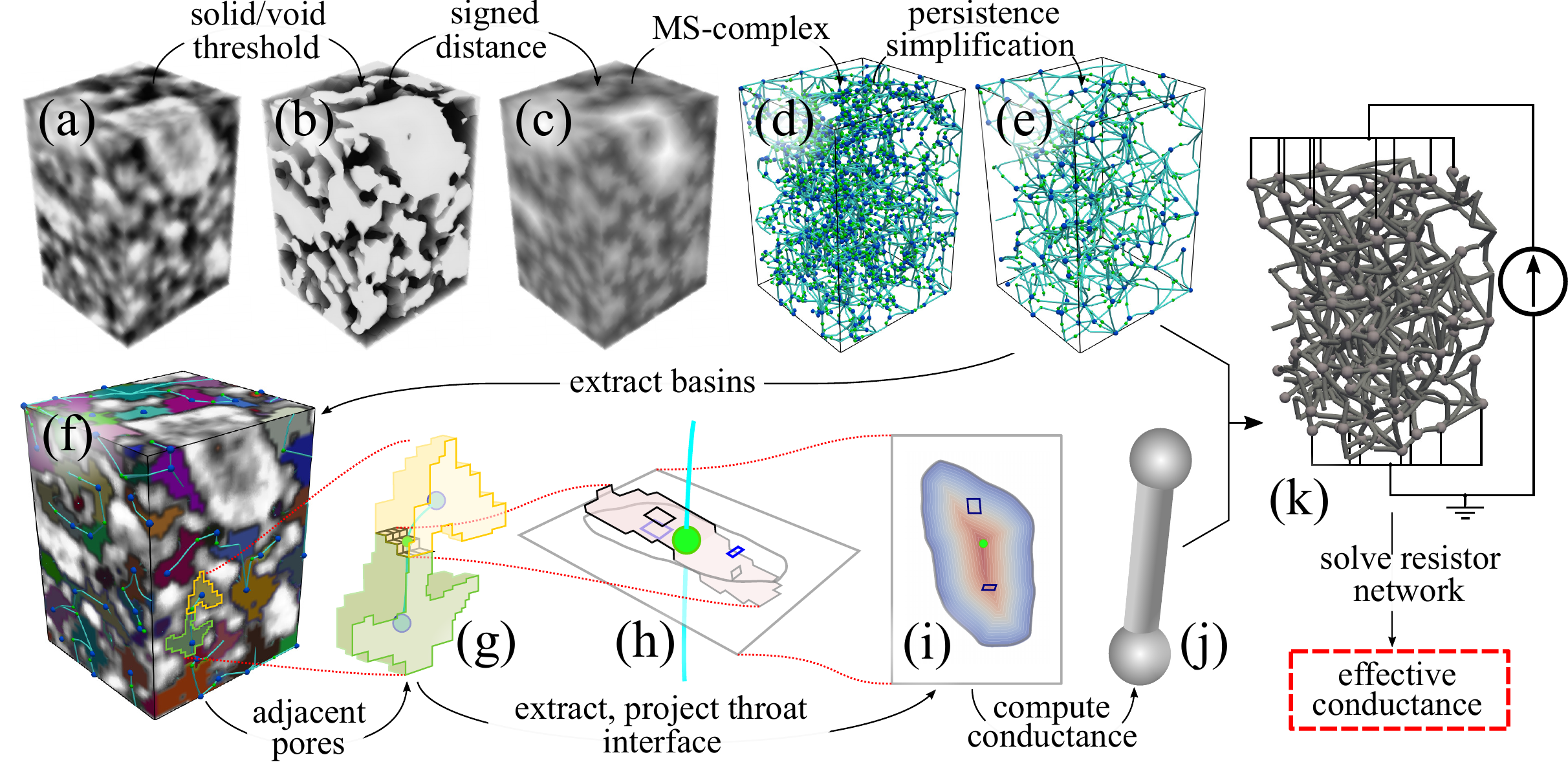}
 \caption {An overview of the image processing, topology computation, and pore network construction. Rectangular regions are selected from the micro-CT (a). The image is partitioned into solid/void with an intensity threshold picked to reproduce the experimentally measured material porosity (b). A signed distance function from the solid/void interface is computed (c). We use MSCEER~\cite{MSCEER} to compute the Morse-Smale complex, encoding the topology of the distance function. The 1-skeleton of the void space is shown in (d): minima (blue spheres), 1-saddles (green spheres), and the arcs connecting them (teal tubes). The over-segmented MS-complex is simplified up to a persistence threshold (e). Basins are computed for the minima that remain after simplification, and form \textit{pores} (f). Each pore is displayed with a random color. The 1-saddles identify the interfaces between pores, called throats (g). The conductance of a throat between basins is dependent on the cross-section shape and area. The boundary quadrilaterals between pores are projected onto a plane normal to the path (h). The fluid velocity normal to the plane through each projected area patch, is computed and summed over the throat (i). The conductivity of a pipe (pore-throat-pore) is derived from the throat conductivity and the length of the path (j). Each pipe is added to a resistor network, with pores on the inflow attached to a virtual current source, and pores on the outflow attached to a virtual ground (k). The network is solved to get the material conductance as well as pressure at each pore and flow through each throat.} 
 \label{fig:approach}
\end{figure*}

\section{Related Work}
\label{sec_relatedwork}

Pore Network Modeling has received a great deal of interest, as it simplifies the estimation of volume flow rate and flow-permeable surface area, in turn, leading to higher throughput analysis of materials, better understanding of the underlying flow physics, and deeper insight into material characteristics. In contrast to solving the elliptic Navier-Stokes equations on the void space using FEM, immersed boundary methods, or lattice-Boltzmann methods, solving the PNM can be done with simple Gauss-Seidel over relaxation~\cite{Hu2012} and conservation of mass, or a solution to a system of linear equations with nodal analysis.
%~\cite{}\todo{ solver methods}. 
Sufian \textit{et al.}~\cite{Dini2019} showed that the PNMs accurately predict the drops in pressure between pores when compared to numerical solutions of the Navier-Stokes equations. Gackiewicz \textit{et al.}~\cite{Cheng2021} also showed that PNMs computed with the maximum-ball and Delaunay method agreed with FEM solutions to the Navier-Stokes equations for sphere packed materials. 

PNM computation falls mainly into four categories: Delaunay tesselation of grain centers, maximum-inscribed ball transform of the void space, medial-axis transform, and watershed-based segmentation. In each case, \textit{pores} are identified, and the interface of adjacent, incident pores is the \textit{throat}, together forming the nodes and pipes of a PNM. For materials formed of packed spheres, a vertex is placed at each particle center, and the Delaunay tesselation~\cite{Lee2004TwoAF} is computed. Tetrahedra are merged using various criteria to form the pores~\cite{Shire2016}, and their faces identify the interfaces, the \textit{throats}, between the pores. A weighted Delaunay triangulation can account for variable sphere sizes~\cite{Tordesillas2017}. Combining the Delaunay tesselation with its dual, the Voronoi tesselation can be used to merge pores, estimate flow rates through throats~\cite{Hu2012}. The Delaunay-based methods work well for packed spheres, but do not generalize to realistic particle/void shapes. 

Maximum-inscribed ball methods simulate the morphological opening of a spherical structuring element~\cite{Jeng2019}, marking each voxel of the void space with the maximum radius ball that fits in the void and covers the voxel~\cite{Dong2009}. Overlapping balls define an ancestry relation for merging regions into pores, with common children becoming the throats. The pores identified can be combined with a medial axis to divide throats into half-throats for a more localized and accurate computation of conductance~\cite{Raeini2017}. The medial-axis transform of the void space was used to find the 2D skeleton for fracture networks~\cite{Li2017}.
%\todo{improve this}.
Several approaches use watershed, or watershed combined with the distance transform to identify pores and throats. Gostick~\cite{Gostick2017} computed the distance transform on the void space, applied a maximum filter to test for maxima, and computed the marker-controlled watershed to segment the void space into pores. Extraneous peaks were trimmed to avoid over-segmentation, and overlap of dilated pores recovered the connectivity.

\paragraph{Topological analysis}
\label{sec:MSC}
Direct study of the topology of scalar functions has led to highly effective approaches for segmentation or encoding domain-specific features of interest. Publicly available tools, based on efficient algorithms~\cite{Robins11, Gyulassy2019} for their computation, such as the Topology Toolkit~\cite{ttk} and MSCEER~\cite{MSCEER}, provide accessibility to the broader community. The Morse-Smale complex, in particular, encodes the gradient flow features of a scalar function, including those of interest in porous media: minima, basins, valley lines, 1-saddles, and the interfaces between basins. These same topological features form the basis for the analysis of: electronic potentials in quantum chemistry~\cite{TopoMS, Olejniczak2020}, the filamentary and dark matter structure in cosmology~\cite{Sousbie2011}; the formation of bubbles in mixing fluids~\cite{Laney06vis}; the core structure of open cell foams~\cite{Gyulassy07btvcg, Petruzza2019}; lithium diffusion pathways~\cite{GKLW16, Gylassy2017}; and many others~\cite{bock_topoangler_2018,
rieck_clique_2018, Bremer09tvcg}. For extracting the pore structure for porous materials, Homberg \textit{et al.}~\cite{Ulrike2012, Ulrike2014} described computing the pores and throats of porous materials in terms of the Morse complex of the distance function. Pores were identified as the descending 3-manifolds from maxima, with the connectivity given by ascending 1-manifolds from 2-saddle. We use an equivalent definition in our approach, adapted to our convention of using negative distance values in the void space and positive in the solid. Furthermore, they studied the pore and constriction size distribution with different merging criterion, which we extend beyond in developing models to compute the conductance of each constriction, transform the pore graph into an equivalent electrical resistor network, find the flow and pressure drop between pores, and compute the effective conductance of the porous material. Ushizima \textit{et al.}~\cite{Gunther} proposed a topolgoy-based pore network model, using the Reeb graph of the height function applied to the void space, to study fluid flow for carbon sequestration. This approach assumes the level set of the height function to be perpendicular to the fluid flow direction, which will not accurately model the flow in the case where the orientation of the flow path is not aligned with the axial direction of the dataset. Furthermore, the use of the Reeb graph necessitated defining an external function on the void space (the height function), which is avoided when using Morse complexes. 

A related topological approach, Percolation theory, studies the connectivity of an infinite network in terms of the size and extent of the largest connected component, as vertices or edges are included/excluded from the network \cite{broadbent_hammersley_1957, Berkowitz1998PercolationTA}. A discontinuity in the resulting \textit{percolation function} provides a threshold that describes an intrinsic porosity property of the material~\cite{koeppfriederici19}. It has been adapted to finite domains and large material images~\cite{friedericikoepp19}.

\section{Approach}
\label{sec_approach}

Our Pore Network Model is built using the topology of the signed distance function to the interface between the solid material and the void. The topological model inherently provides a means of clustering pores through persistence simplification~\cite{edelsbrunner}. It encodes not only valley-lines as paths but also the spatial decomposition of the voids into basins, similar to the Watershed transform. Conductance values are computed for each interface between regions connected by a saddle and combined into a resistor network, which is then solved to obtain the effective conductance of the sample and all the internal flows. We compute material properties such as volumetric flow rate by converting from electrical flow to fluid flow and flow-permeable surface area by examining the current flow through the pores.       

\subsection{Image Processing}
\label{sec:improc}
Given a 3D micro-CT image, we first select a rectangular region from the interior of the typically cylindrical scan (\autoref{fig:approach} (a)) to avoid potentially confounding effects of the imaged container on the measured material properties. We then compute a two-phase segmentation into solid and void, using the measured weight and grain density for each sample provided by our collaborators (\autoref{fig:approach} (b)). An intensity threshold is chosen such that the fraction of foreground to background voxels is $1 - \epsilon$, with $\epsilon$ chosen such that the total volume of the foreground material multiplied by the known density of the grains equals the measured weight. A signed distance field (negative in the void space, positive in the solid) is computed by inserting points from the isosurface at the chosen intensity threshold into a $k$-d tree and computing the nearest interface point for each voxel, (Figure~\ref{fig:approach} (c)). Note that this approach discretizes the distance field by sampling it onto the voxels of a grid -- for small throat diameters, the image artifacts created by the stair-case like sampling  distorts both the distance values and the geometry of the void space. We mitigate these challenges by evaluating the distance field with respect to a two-fold refined grid in each axis. As discussed below, this sampling density has been sufficient for all subsequent processing and analysis steps. 

% Get a two-phase representation 

% For simple/simulated examples, binary already know

% For micro-CT, simply apply a threshold that creates the known material porosity. 

% Distance field

% Dense set of points on material/void boundary, 

% Signed distance

\subsection{Topological decomposition}
\label{sec:topoanal}
Formally, the Morse-Smale Complex (MSC) is defined as follows. Given a compact $d$-manifold $\Mspace$, a scalar function $f: \Mspace \rightarrow \Rspace$ is a \textit{Morse} function if its \textit{critical points} are non-degenerate and have distinct values. A
critical point occurs where the gradient vanishes, $\nabla f = 0$, and is non-degenerate if its Hessian is non-singular. 
For Morse functions,
the neighborhood of a critical point $p$ takes on a quadratic form, and can be written as
$f_p = \pm x_1^2 \pm
x_2^2 \dots \pm x_d^2$, where the number of minus signs in this equation defines
the \textit{index} of criticality. 
For volumetric functions, \textit{minima} are
index-0, \textit{1-saddles} are index-1, \textit{ 2-saddles} are index-2, and \textit{maxima} are index-3. Here we are particularly interested in the \emph{basins} around the minima and the 1-skeleton connecting neighboring basins through their 1-saddles.
%Basins are defined as the collection of voxels whose descending integral lines end in the respective minimum and the 1-skeleton arcs are formed by the steepest descending lines from the 1-saddles to the minima. 
For a more detailed description of the entire Morse-Smale complex, we refer the reader 
to Gyulassy et al.~\cite{gyulassy2012direct}.

We compute the MSC of the signed distance field using an efficient approach based on discrete Morse theory~\cite{MSCEER}, using Robins \textit{et al.} steepest-descent algorithm~\cite{Robins11}. This approach creates a data structure that can extract: the critical points and arcs connecting them, forming the 1-skeleton, whose paths follow steepest ascent and descent through the 6-neighbor connectivity of voxels; and, basins, the collection of voxels that form the terminus of steepest descending paths for each local minimum. This data structure~\cite{gyulassy2012direct} can provide the 1-skeleton and basins for any threshold of \textit{persistence}, a metric that removes topological features ordered by increasing differences in the function value. The unsimplified 1-skeleton is extracted (Figure~\ref{fig:approach}(d)) and simplified to an absolute threshold of 1.4 (Figure~\ref{fig:approach}(e)). The 1-skeleton is smoothed both for visualization and to better estimate distances between pores. The persistence threshold removes topological features that exist solely due to sampling the distance function onto a grid and using 6-neighbor connectivity between voxels. Furthermore, it merges the minima and hence basins that have less than 1.4 voxel difference in their distance to the solid-void boundary. Note that in Section~\ref{sec:topo_analysis}, we refined the sampling of the distance function -- increasing the ability of persistence simplification to remove discretization artifacts \textit{without} removing features due to subtle variations in the void structure diameter. 
In section~\ref{sec:topo_pers}, we show that ultimately the analysis results are not sensitive to this threshold, as \textit{all} imaged material conductivity reacts similarly to variations in it. 

% Topology compute (MSCEER, ms complex, basins, 0-1 arcs, PERSISTENCE) (Figure: tiny cube 
% of 4x mag 100 Segmented Basins below zero, grains semi-transparent, Line network, Line 
% network + interfaces) 4 images

% Use discrete morse theoretic approach from MSCEER

% Builds 1-skeleton for interactive simplification

% Based on exploration & reasoning, we want the  lowest persistence threshold that 
% removes artifacts

% Come from binarization (jagged edges)

% Inexact distance transform (could be off by ½ voxel)

% Sampling distance onto a grid

% 6-neighbor connectivity of the discrete gradient

% Empirically, means persistence = 1.2

% For materials with small holes, this means we will have to increase resolution before 
% step 2a - empirically, needed 2x increase for examples

% The basins of the simplified MS-complex are constructed

% Topology network to PNM model
\subsection{Pore Network Model}
The simplified decomposition into basins and concurrent simplified arcs of the MS-complex (Figure~\ref{fig:approach}(f)) are converted into pores and throats of the pore network model. A pore is created for each minimum and its associated basin, and is assigned the voxels in the basin having a distance value less than 0 (indicating they are part of the void space).
% The voxels with dist < 0 are given basin minimum id, = pores
Each 1-saddle having a distance value less than 0 and connecting 2 distinct minima in the 1-skeleton of the MS-complex becomes a throat. 
In the case that distinct 1-saddles connect the same pair of minima, the lowest-valued one (and consequently, the least restrictive) is selected to represent the throat. The geometry of the throat is formed by the quadrilateral faces that separate voxels from different pores (figure~\ref{fig:approach}(g)). Each quadrilateral is used as an area element to compute the conductance in a cross-section perpendicular to the flow. The projected area of each unit quadrilateral face is $a(q) = |n_q \cdot n_p|$, where $q$ is the quadrilateral, $n_q$ its normal, and $n_p$ the normalized direction of flow, estimated using the arc of the MS complex (figure~\ref{fig:approach}(h)). 

\paragraph{Computed conductance of a throat} The smoothed path through the 1-saddle provides a center-line for the virtual ``pipe'' that connects two pores. We make simplifying assumptions for computing the conductance of this pipe: (1) that the laminar flow is fully-developed along the length of the pipe; (2) the velocity profile is quadratic, and (3) that the plane normal to the pipe and intersecting the 1-saddle is a representative cross-section. For perfect cylinders, our assumptions are equivalent to that in \autoref{eqn:H-P}. The 1-saddle occurs as the point of greatest absolute distance $R$ on this plane. In viscous fluid flow, the pressure difference at the ends of each pipe is balanced by the shear stress due to viscosity, where fluid sticks to the pipe's walls and its velocity, $u = 0$. The instantaneous velocity at any point on the cross-section of a cylindrical pipe is given by
\begin{equation} \label{eqn:velprof}
    u(r) = \frac{\Delta p}{4\eta L}[R^2 - r^2],
\end{equation}
depending on the pressure drop $\Delta p$, viscosity $\eta$, and the length of the pipe $L$, which act as constants over the pipe, as well as the distance from the center of the pipe, r. In the computation of the volume flow rate $Q_{pipe}$ through the pipe, $u(r)$ is integrated over the cross-sectional area, which we evaluate as the sum over area elements projected onto the plane,
\begin{equation} \label{eqn:sum}
    Q_{pipe} = \frac{\Delta p}{4\eta L}\sum_q a(q) (R^2 - (R-d(q))^2),
\end{equation}
where $d(q)$ is the absolute value of the distance function evaluated at the center of the quad (figure~\ref{fig:approach}(i)). Note that for non-circular cross-sections, the velocity $u(r)$ is slightly over-estimated. The conductivity of the pipe $C_{pipe} = Q_{pipe} / \Delta p$, factors out the yet-unknown drop in pressure. 

% We evaluate  
% In this pipe, the the direction flow will take, and each quadrilateral is projected onto a plane normal to this path and intersecting the 1-saddle (figure~\ref{fig:approach}(h)).
% PNM model conductance for each throat 

% Fully developed flow in circular pipes Flow velocity across cross section of a circular 
% pipe

% Give equation: 

% Say velocity is proportional to the pressure drop, the radius, inversely proportional 
% to the viscosity and length

% Closer to the wall the lower the velocity

% The $(R^2 - r^2)$ controls the conductivity of the cross section

% Delta p will be found later

% Conductivity of a pipe/throat - take out the delta p

% Integral over cross section of velocity

% 4mew, L are constants w.r.t. A pipe

% Simplified model: 

% fully developed laminar flow along entire pipe

% $Area = sum_{area elements}$ area(a) dot normal - explain area elements are the quads 
% that separate voxels with different ids, project onto plane

% Use $sum_{area elements}$ $(R^2- r^2)$, where R is radius of 1-saddle, r is R - distance() 

\subsection{PNM solved with Resistive Network}

Having computed a conductance for each throat between pores by modeling them as pipes, we build a pore network and convert it to a resistor network so that the voltage and current can be solved from a simple linear system of equations. The conservation of mass in Navier-Stokes equation is mirrored by Kirchhoff’s Current Law, that the sum of currents coming into a node equals the sum leaving it. Each pore in the pore network becomes a node $n_i$ for nodal analysis. In addition, we create a virtual node to act as the current source and another to act as the virtual sink or the ground. Each throat becomes a resistor $r_{i,j}$, connecting nodes $n_i$ and $n_j$, with resistance $1/C_{pipe}$, and the nodes on the inflow and outflow surfaces of the sample are connected to the virtual source and sink respectively (figure~\ref{fig:approach}(k)).

The network is solved by adding each resistor $r_{i,j}$ to an admittance matrix $M$, contributing a positive admittance $1/r_{i,j}$ to the diagonal elements $M_{i,i}$ and $M_{j,j}$, and a negative admittance to off-diagonals $M_{i,j}$ and $M_{j,i}$. The system $Mx=b$ is solved, where $x$ is the vector of unknown voltages, and the current vector $b$ is initialized with $b_{source} = -1$ and zeroes for the remaining nodes. The system is solved with a generalized minimal residual method for sparse matrices (GMRES~\cite{BLAS}). After solving, the vector $x$ gives the voltage for each node in the network. The directional current, and hence volumetric flow rate, for any throat, is simply the voltage drop divided by the resistance, $Q_{i,j} = (x_i - x_j) / r_{i,j}$, with the sign of $Q_{i,j}$ determining the direction of flow between $n_i$ and $n_j$. The source-to-sink voltage drop given the input unit current gives the effective resistance of the block of material, which is converted back to flow conductance, $C_{e} = 1/ (x_{source} - x_{sink})$. 

% Pore = node for nodal analysis, virtual source, virtual sink,

% Each throat becomes a resistor, (1/conductance )

% Apply unit current

% Incidence matrix $Mx=b$ where 

% M is a sparse matrix of admittance values

% $m_i,j$ with admittance value between node i, j, 

% For each edge (i,j) in the pore network, we add a positive admittance value to the 
% diagonal (i,i and j,j) and a negative admittance value to non-diagonal (i,j and j,i).

% x is vector of voltages we want to solve, 

% b = [-1,0,0,..] initial current vector 

% Use GMRES to solve for voltage at each node

% Final b gives current, final x gives voltage at the pore, which is transformed into 
% voltage drop across a pipe and current flowing through the pipe.

% Effective resistance of the material = source-to-sink voltage drop
\subsection{Computing material properties}

The resistive network solution and recasting in terms of fluid flow provides several insights into the porous materials, from aggregate properties over the sample to per-pore and per-throat properties. For each pore, the pressure, total in/out flow, the geometric embedding of its voxels, and its portion of the total material surface area can be of interest. For each throat, the pressure drop, volumetric flow rate, orientation, path, length, radius, and aspect ratio can be of interest. 

Aggregate behavior of the material sample, such as volumetric flow rate, flow-permeable surface, and tortuosity factor, can also be computed from the solution of the resistive network. The volume flow rate through the sample is calculated for any $\Delta p$ by multiplying by the computed effective conductance $Q_e = C_e \Delta p$. The sum of triangle areas of the isosurface at the solid/void interface threshold gives the total area; sampling the pore id, and thus volume flow rate, at the center of each triangle allows discarding triangles whose computed flow fall below a user-selected epsilon threshold to produce the aggregate flow-permeable surface area. The tortuosity measure can be computed by averaging the shortest paths connecting each pore connected to the network source, to the network sink, and vice versa. An alternate calculation for tortuosity ``counts'' the distance all electrons traveled by computing the sum of path lengths weighted by the current they carry.  

Finally, the CK equation provides a mechanism for converting volumetric flow rate to the flow-permeable surface area and back again, meaning our computation of each via resistive network and the material isosurface give two complementary methods of computing each.

\subsection {Visual Verification}
The complexity of the overall computational pipeline required each step to be verified visually. Most steps were visualized with an OpenGL viewer built on top of MSCEER~\cite{MSCEER}, as well as output sent to ImageJ and ParaView. The custom visualization of flow paths combined with visualization of the total in/outflow into basins (as in ~\autoref{fig:teaser}, right) was instrumental in building credibility that the vast majority of the void space admits flow, i.e., that dead-ends do not play a significant role.

% phasearea are directly 
% Material properties:

% Compute wetted surface area from isosurface:

% Make a binary volume, where v = 1 if basin has inflow

% 0 if basin has no inflow

% Compute isosurface material/void boundary (through threshold)

% Sample volume at centroids of triangles

% Do weighted sum of triangle areas with 1/0 weights

% Calculate effective flow rate through substitution with Carman-Kozeny equation, using 
% the 5.3 shape factor the scientists use...

% Direct computation of effective flow rate from computed effective conductivity:

% 1/effective resistance = conductivity of material - used to get effective flow rate by 
% $Q= C_e / delta(P).$

% Calculation of wetted surface area from effective flow rate

% Use c) + Carman-Kozeny to get wetted surface area

\section{Validation}
\label{sec_validation}
We validate our PNM using simple configurations where the conductance of the void space can be calculated using the Hagen-Poiseuille equation 
\begin{equation} \label{eqn:H-P}
     C = \frac{\pi r^4}{8 \eta L}
\end{equation}
where $r$ is the radius of a cylindrical pipe, $L$ its length, and $\eta$ is the dynamic viscosity.

\begin{figure}
    \begin{subfigure}[h]{0.3\columnwidth}
        \centering 
        \includegraphics[width=\columnwidth]{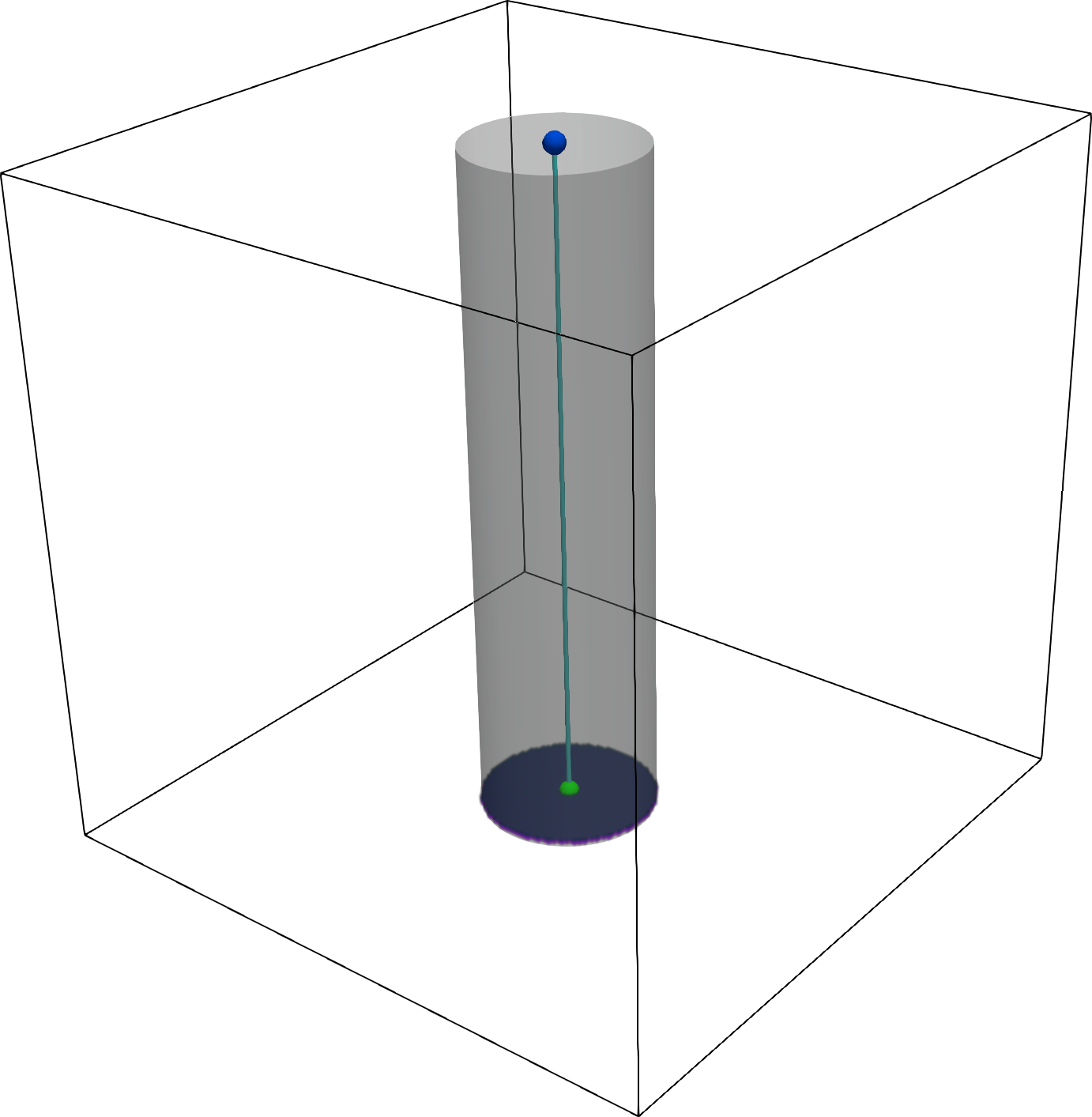}
        \caption{ }
        \label{fig:tube_interface}
    \end{subfigure}
    % \begin{subfigure}[h]{0.5\columnwidth}
    %     \centering 
    %     \includegraphics[trim=600 50 600 150,clip, width=\columnwidth]{figures/tube_verticle_seg.png}
    %     \caption{ }
    %     \label{fig:tube_seg}
    % \end{subfigure}
    \begin{subfigure}[h]{0.3\columnwidth}
        \centering 
        \includegraphics[width=\columnwidth]{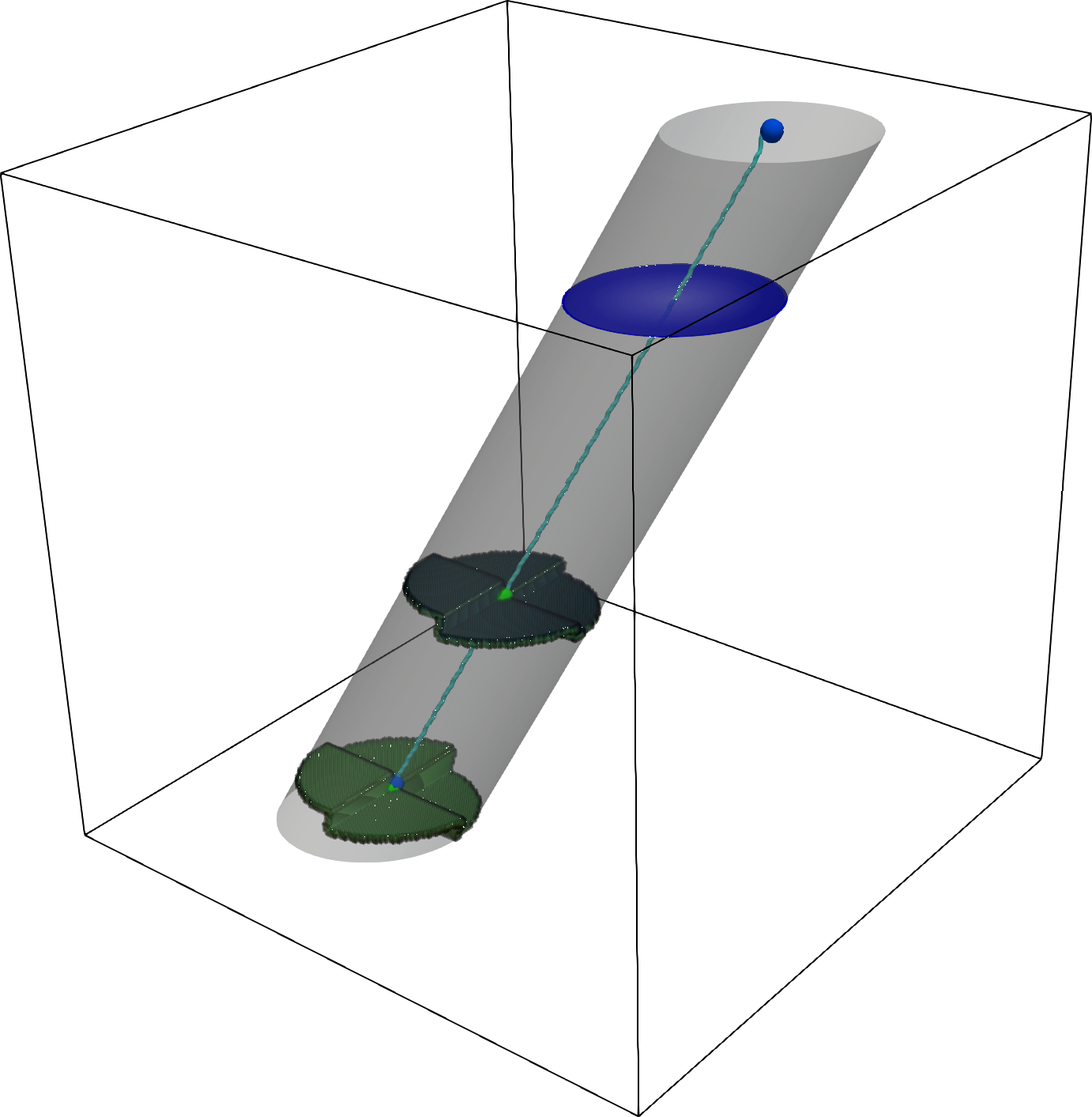}
        \caption{ }
        \label{fig:tube_30_interface}
    \end{subfigure}
    \begin{subfigure}[h]{0.34\columnwidth}
        \centering 
        \includegraphics[width=0.95\columnwidth]{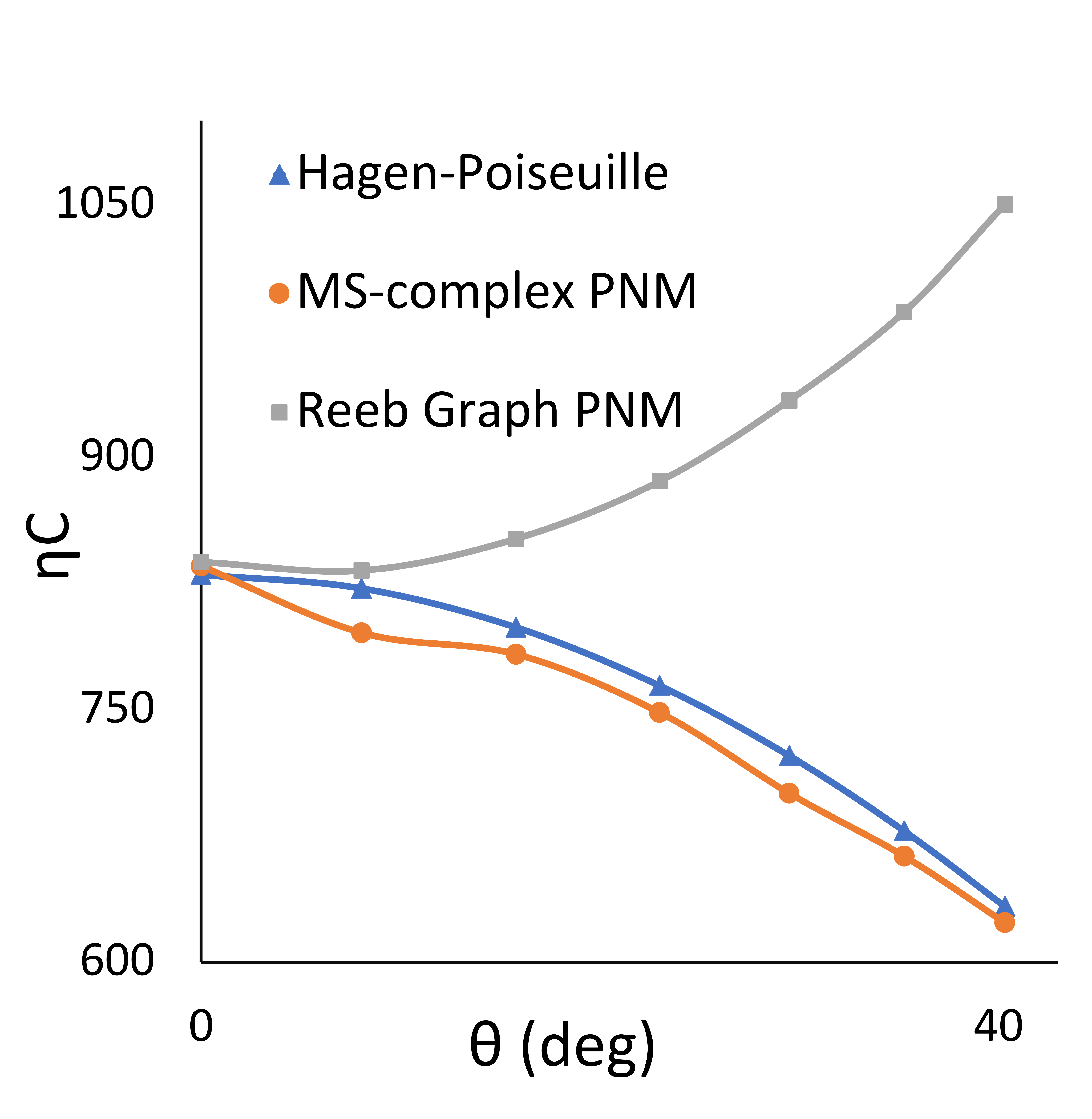}
        \caption{ }
        \label{fig:conductance_comparison}
    \end{subfigure}

    %     \begin{subfigure}[h]{0.5\columnwidth}
    %     \centering 
    %     \includegraphics[trim=600 50 600 150,clip, width=\columnwidth]{figures/tube_29.5_deg_seg.png}
    %     \caption{ }
    %     \label{fig:tube_30_seg}
    % \end{subfigure}
    \caption{The critical points, 1-skeleton of the MS-complex, and pore interfaces for a cylindrical pipe oriented at $\theta= 0$ (a) and $\theta = 29.5$ (b). Note: the blue disk in (b) is a single random slice through the volume, shown here for reference, not to be confused with the pore interface. (c) Conductance computed using MS-complex PNM, Reeb Graph PNM and the Hagen-Poiseuille equation as a function of the pipe angle. The small errors in our PNM with respect to the ground truth are likely attributable to the underestimation of the pipe radius due to digitization of the distance function.}
    \label{fig:pipes}
\end{figure}

Our first simulated dataset models the void space as a single cylindrical pipe, of length $L = \frac{D}{\cos\theta}$, that connects an inlet at the top of the Z plane to an outlet at the bottom of the Z plane (\autoref{fig:pipes}), where $D$ is the depth of the volume, and $\theta$ is the angle between the unit z-direction and the flow direction. For various orientations of the pipe $\theta$, we calculate the effective conductance for air flow, and verify that our model computes the correct pores, throats, pipes, and the cross-section areas. \autoref{fig:tube_interface} shows our PNM for two such orientations, $\theta = 0$, and $\theta = 29.5$ degrees. For both the orientations, our PNM correctly identifies the pore centers (marked as blue spheres), throats (green spheres), the cross-section area of the interface (plane normal to the flow, intersecting the 1-saddle) and the pipe (pore-throat-pore path). Note that the minimum on the top and the bottom Z plane is guaranteed by the stratified boundary handling of the discrete gradient, guaranteeing the existence of a 1-saddle, and the minimum-saddle-minimum path in the interior. For $\theta=0$, our model computes a single pipe; for $\theta=29.5$ degrees, our model computes two pipes due to a local minimum that was not simplified. Notably, this over-segmentation effectively puts two resistors in series in the flow network model and does not affect the computed outcome. The set of quads forming each interface are aligned with the principal directions of the grid -- however, our projection of their area onto a plane normal to the path, when integrating velocities over the interface, yields the correct value. Neither marker-controlled watershed nor Maximum-ball methods can extract this pipe. Medial-axis methods can identify the pipe, albeit with special boundary handling. In \autoref{fig:conductance_comparison}, we compare our computed conductance with the calculated conductance using \autoref{eqn:H-P} and the conductance computed using the topological approach based on the Reeb graph \cite{Gunther}. Even for this simple configuration, prior work fails to compute the correct PNM and hence the conductance and the material permeability. 

\begin{figure}
    \begin{subfigure}[h]{0.3\columnwidth}
        \centering 
        \includegraphics[width=\columnwidth]{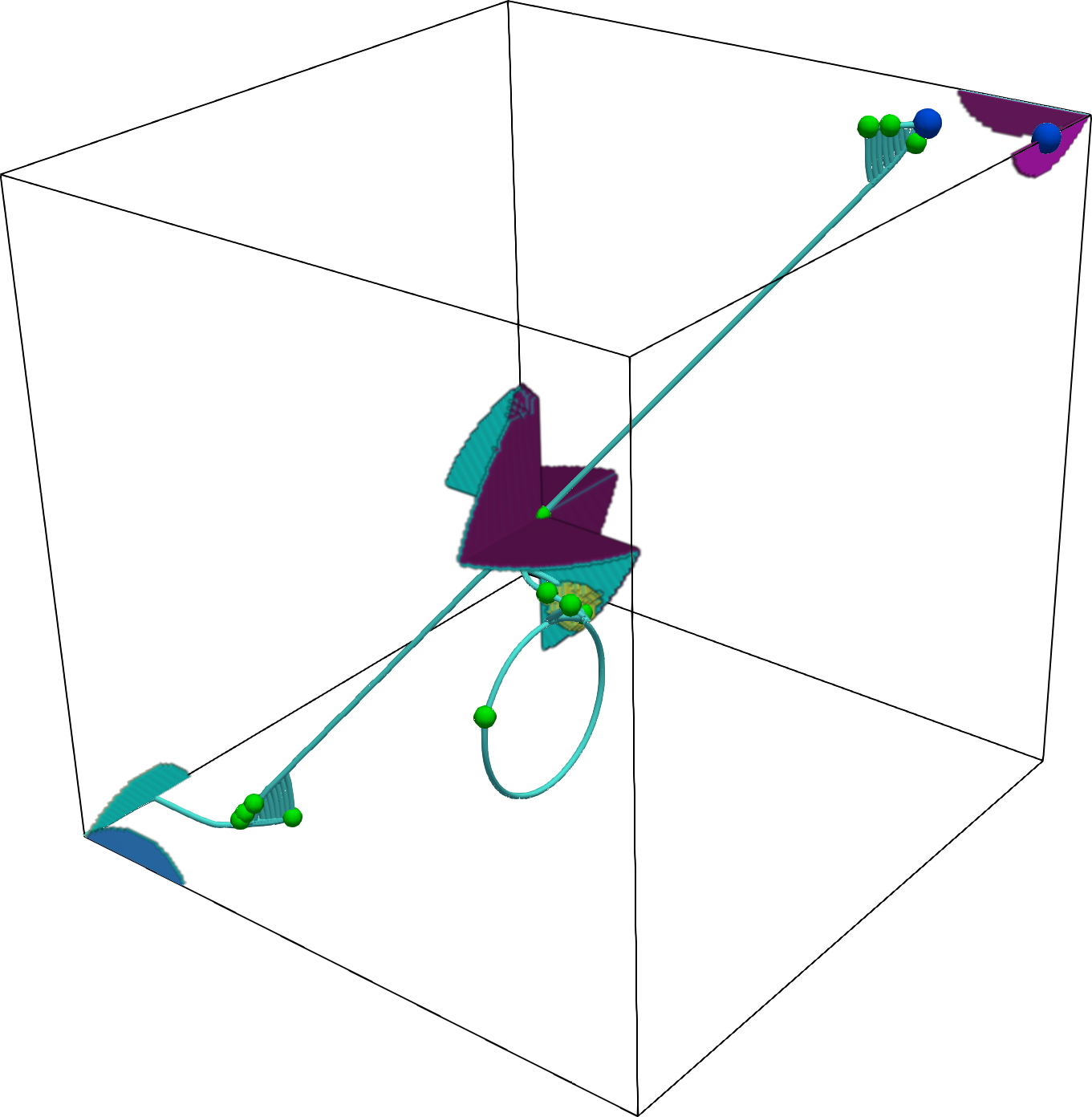}
        \caption { }
        \label{fig:torus_interface}
    \end{subfigure}
    \begin{subfigure}[h]{0.3\columnwidth}
        \centering 
        \includegraphics[width=\columnwidth]{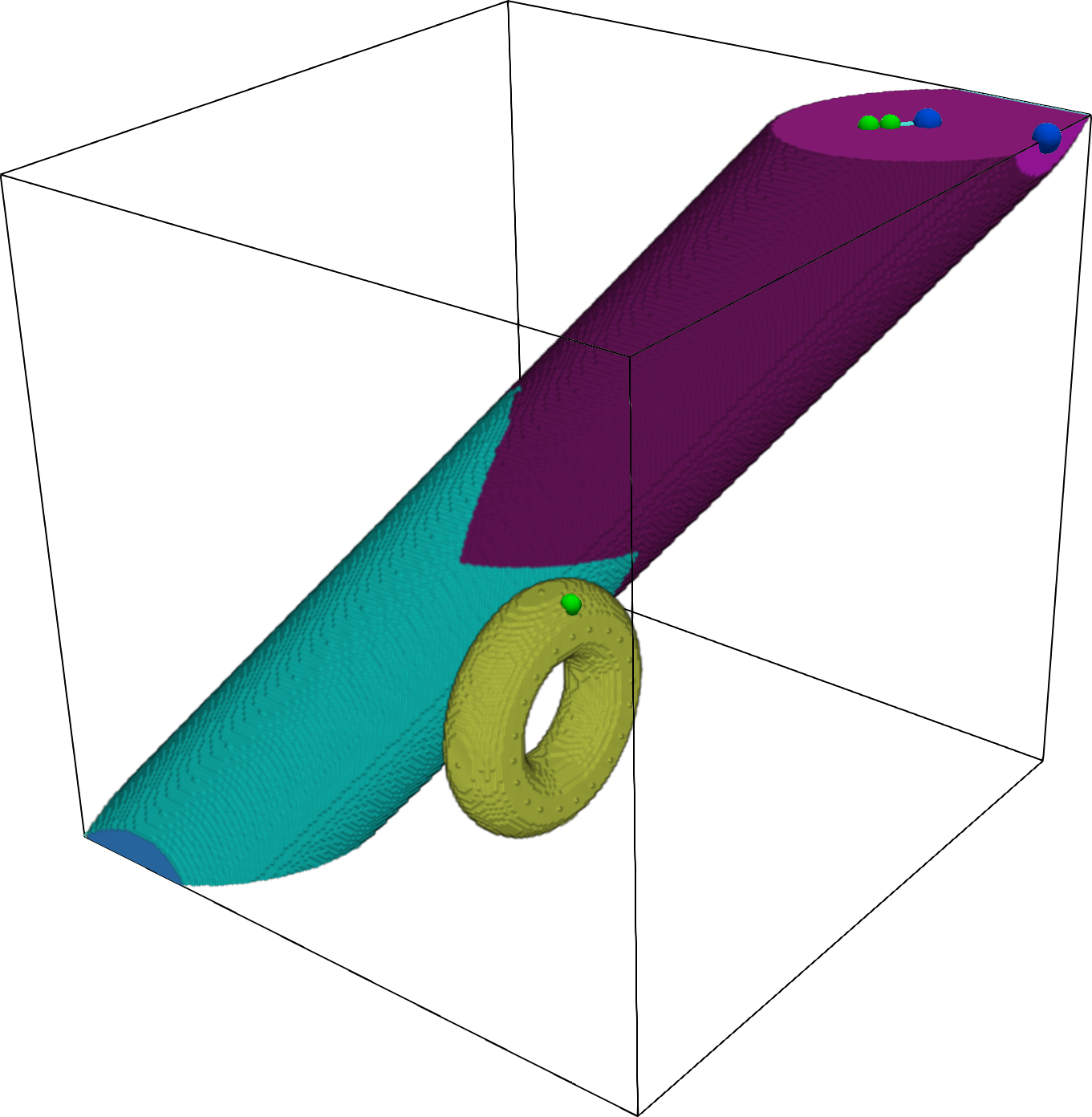}
        \caption { }
        \label{fig:torus_seg}
    \end{subfigure}
    \begin{subfigure}[h]{0.3\columnwidth}
        \centering 
        \includegraphics[width=\columnwidth]{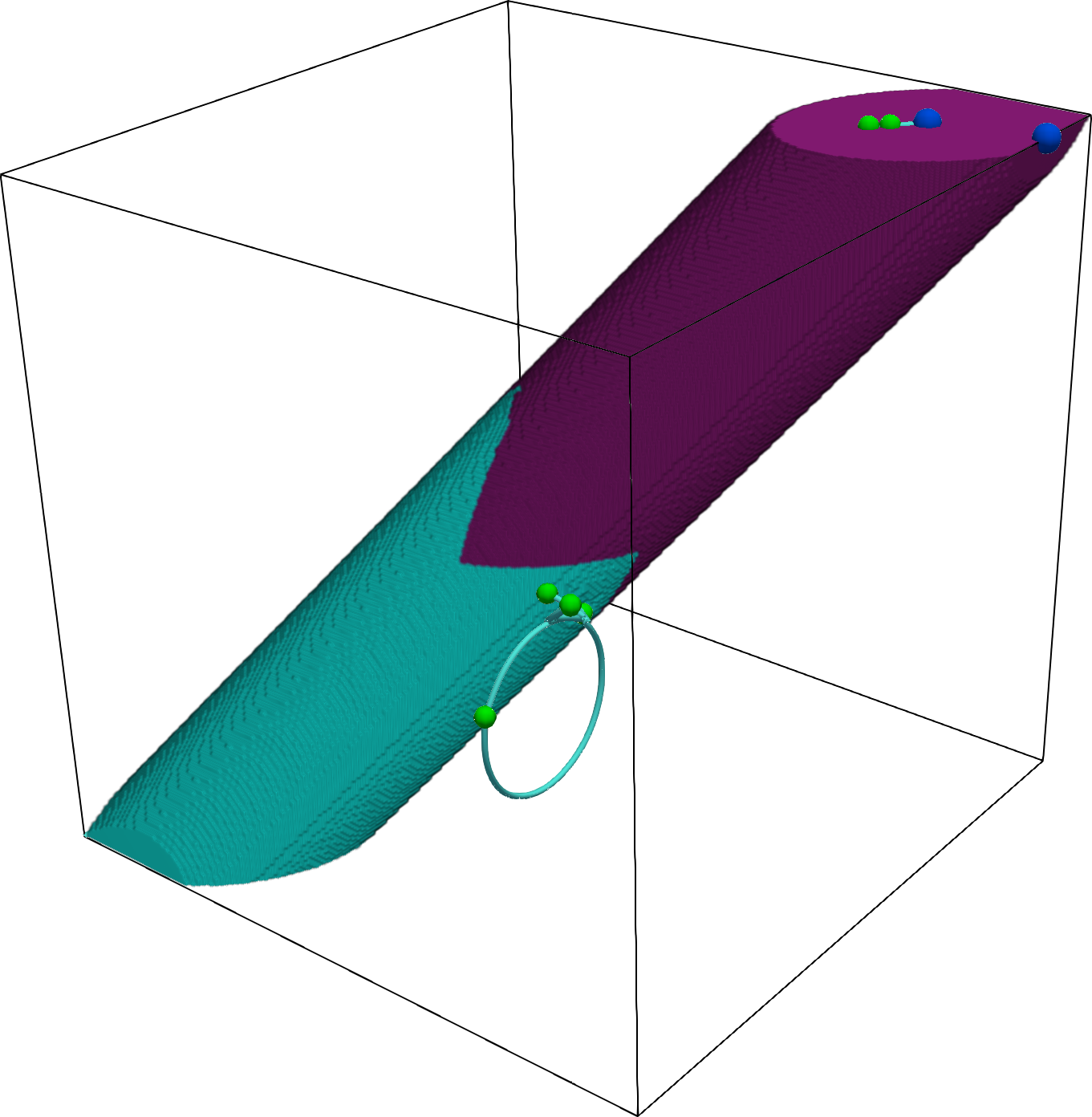}
        \caption { }
        \label{fig:torus_flow}
    \end{subfigure}
    \caption {(a) The critical points and the 1-skeleton of the MS complex, along with cross-section of the interface. (b) Isosurface of the entire pore/void interface colored using pore ids. After solving the PNM, isosurface triangles in non-flow-permeable basins are discarded (c).} \label{fig:torus}
\end{figure}

Our second crafted validation example creates a single cylindrical void, placed at an angle, with a non-simply connected dead-end, carved out by a torus tangentially touching the cylinder (\autoref{fig:torus}). Air/fluid do not flow through a dead-end, and therefore, it is important for a PNM to correctly identify and exclude such regions from the counting towards the virtually measured flow-permeable surface area, which is what the Fisher apparatus measures experimentally. The flow-permeable surface area of the material should be independent of the surface area of the non-simply connected dead-end. \autoref{fig:torus_interface} shows the 1-skeleton of the MS-complex, along with the cross-section of the interface. There are two cross-sections in this visualization. The first one corresponds to the minimum-saddle-minimum arc that connects the minimum at the top and the bottom of the Z plane, and the second one corresponds to the minimum-saddle-minimum arc that connects the minimum on the torus to the minimum at the top of the Z plane. The solution to the resistor network finds no current flow through the dead-end, as expected. \autoref{fig:torus_seg} visualizes the isosurface of the pore space colored using the labels from MSC segmentation before the resistor network solve, and \autoref{fig:torus_flow} shows the isosurface after pieces intersecting non-flow-permeable pores are removed, after the solve.

\begin{figure*}
        \begin{subfigure}[h]{0.35\linewidth}
        \centering 
        \includegraphics[width=0.85\linewidth]{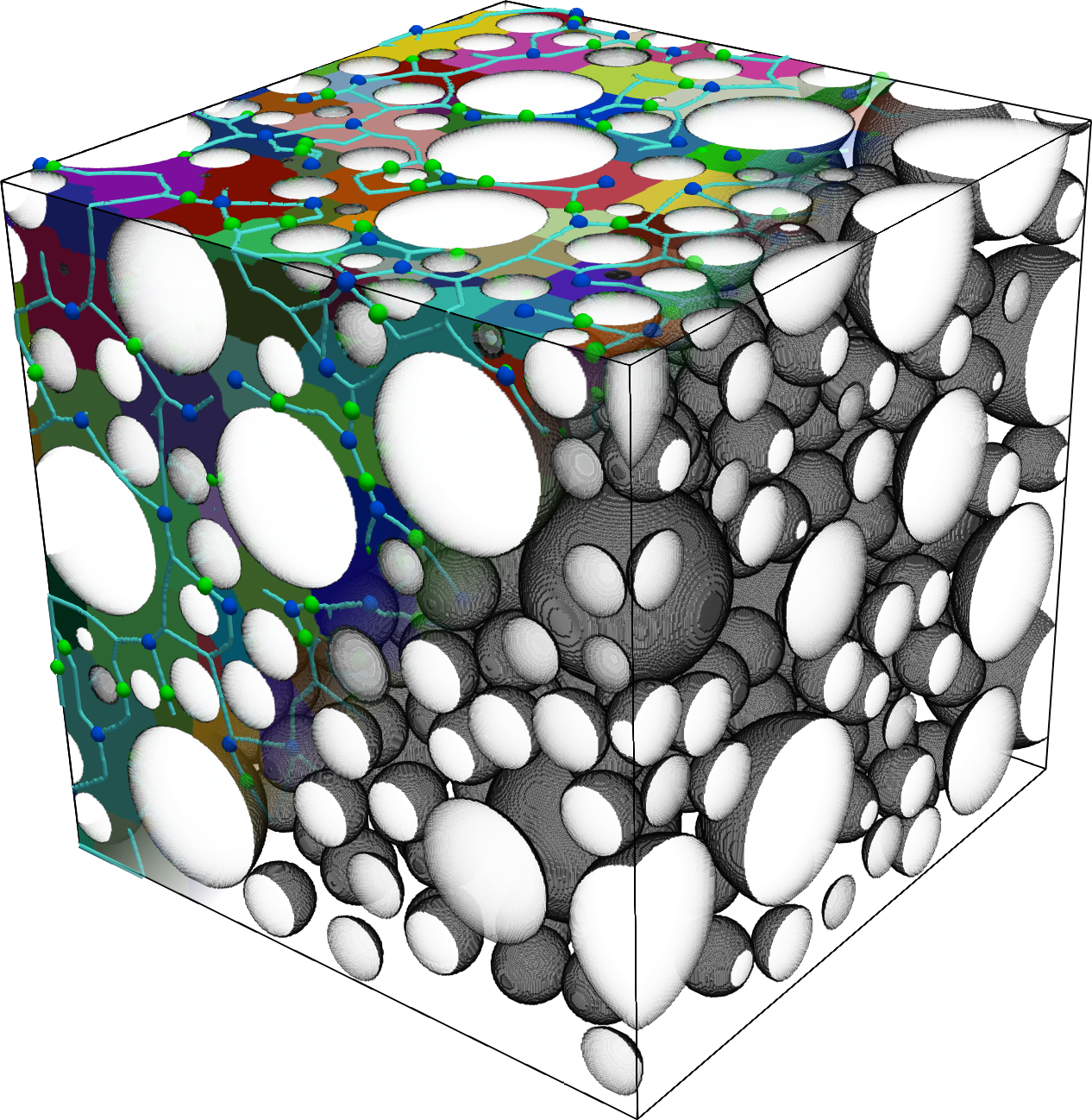}
        \caption { }
        \label{fig:mono-2}
    \end{subfigure}
    \begin{subfigure}[h]{0.35\linewidth}
        \centering 
        \includegraphics[width=0.85\linewidth]{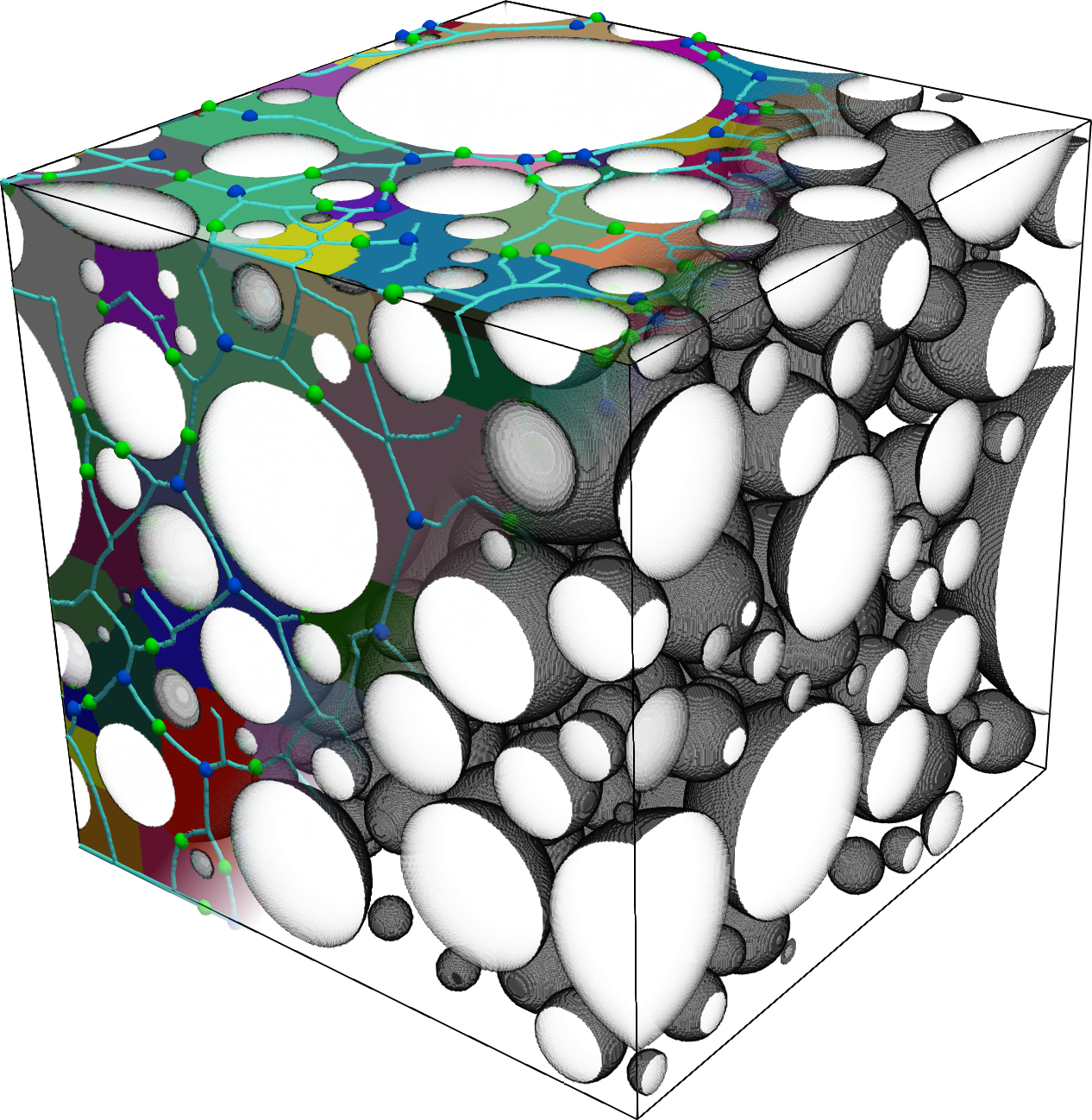}
        \caption { }
        \label{fig:poly}
    \end{subfigure}
        \begin{subfigure}[h]{0.35\linewidth}
        \centering 
        \includegraphics[width=0.85\linewidth]{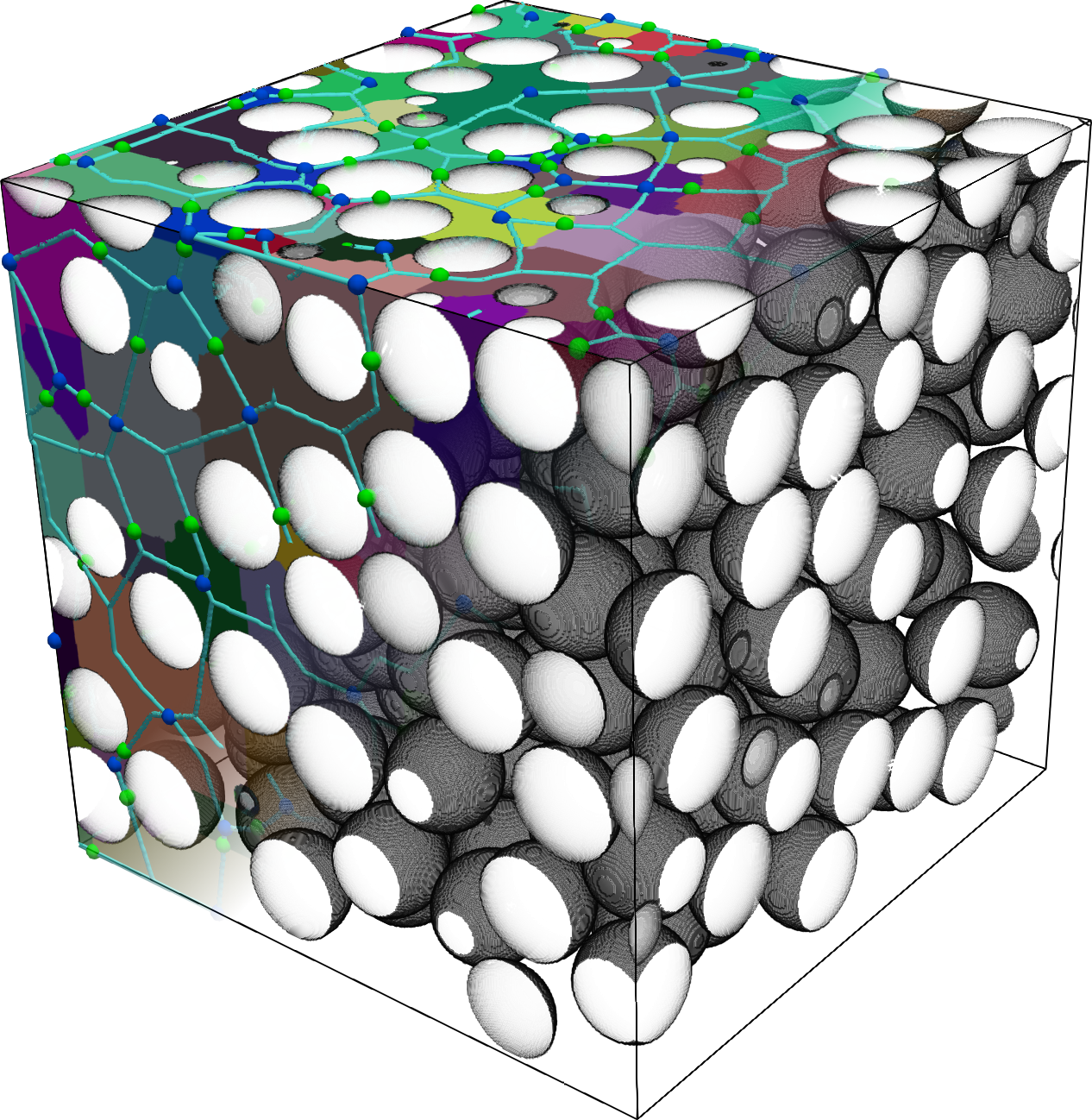}
        \caption { }
        \label{fig:mono-1}
    \end{subfigure}
            \begin{subfigure}[h]{0.35\linewidth}
        \centering 
        \includegraphics[width=0.85\linewidth]{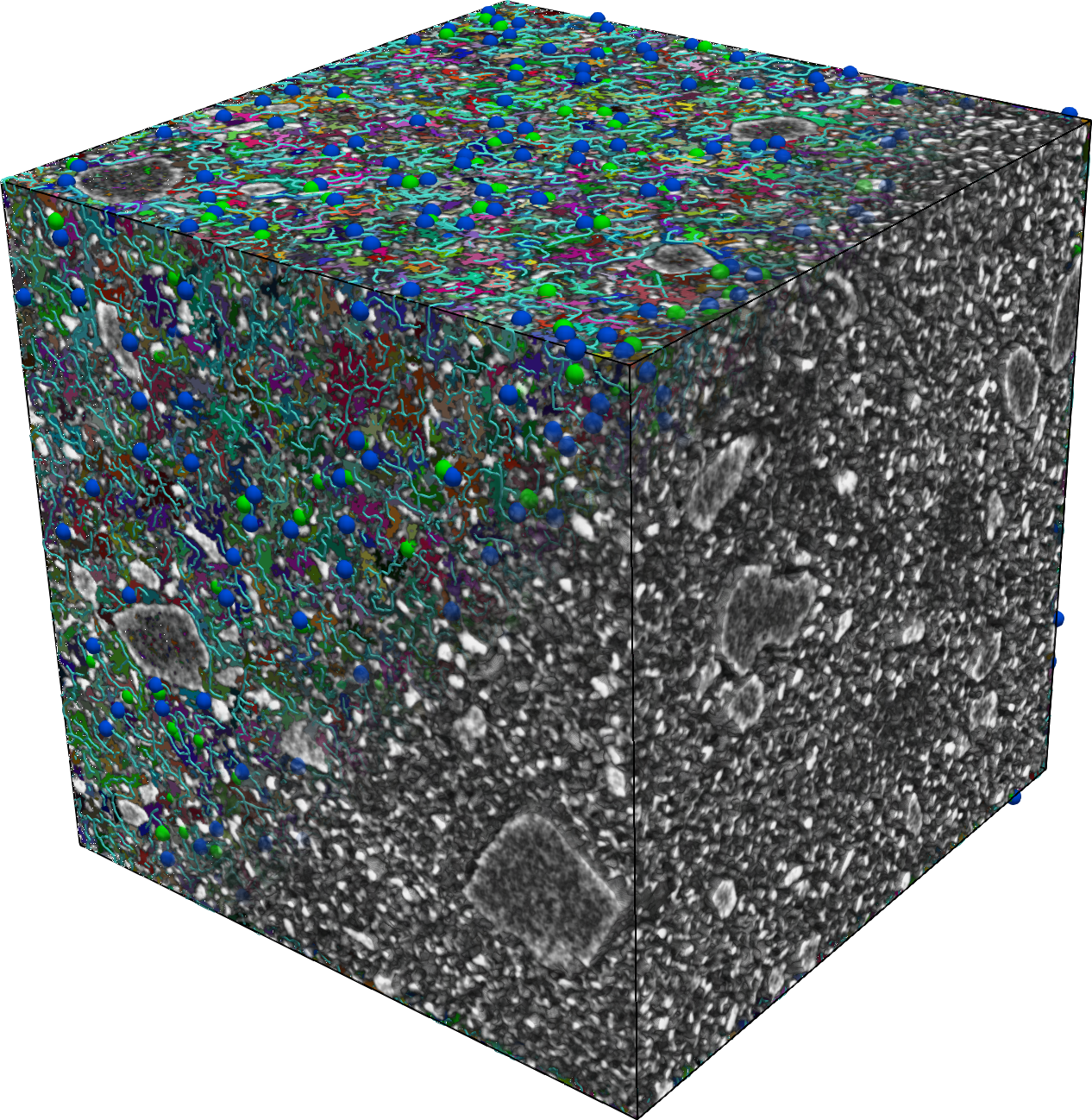}
        \caption { }
        \label{fig:HE_A}
    \end{subfigure}
    \begin{subfigure}[h]{0.35\linewidth}
        \centering 
        \includegraphics[width=0.85\linewidth]{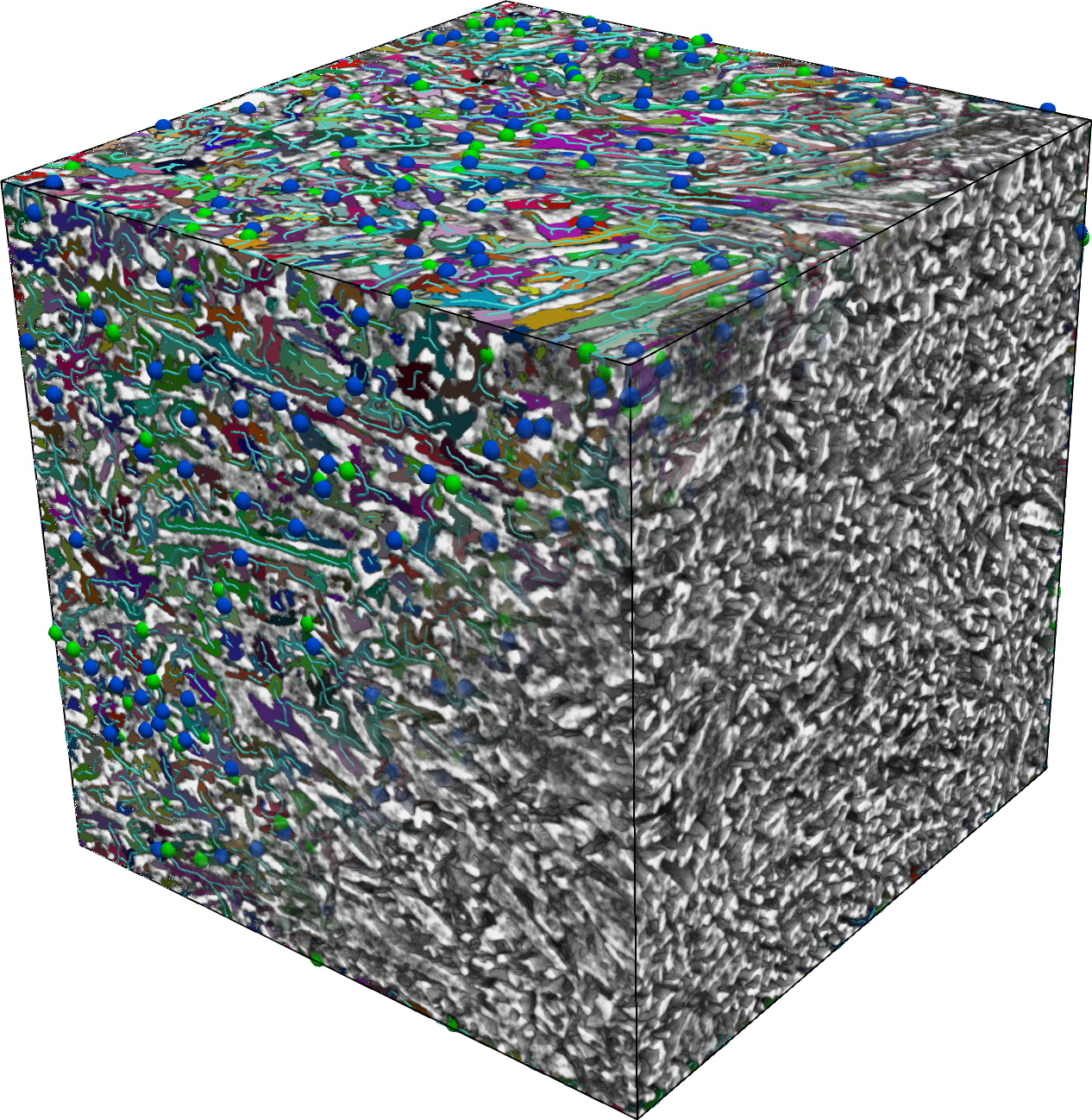}
        \caption { }
        \label{fig:HE_B}
    \end{subfigure}
        \begin{subfigure}[h]{0.35\linewidth}
        \centering 
        \includegraphics[width=0.85\linewidth]{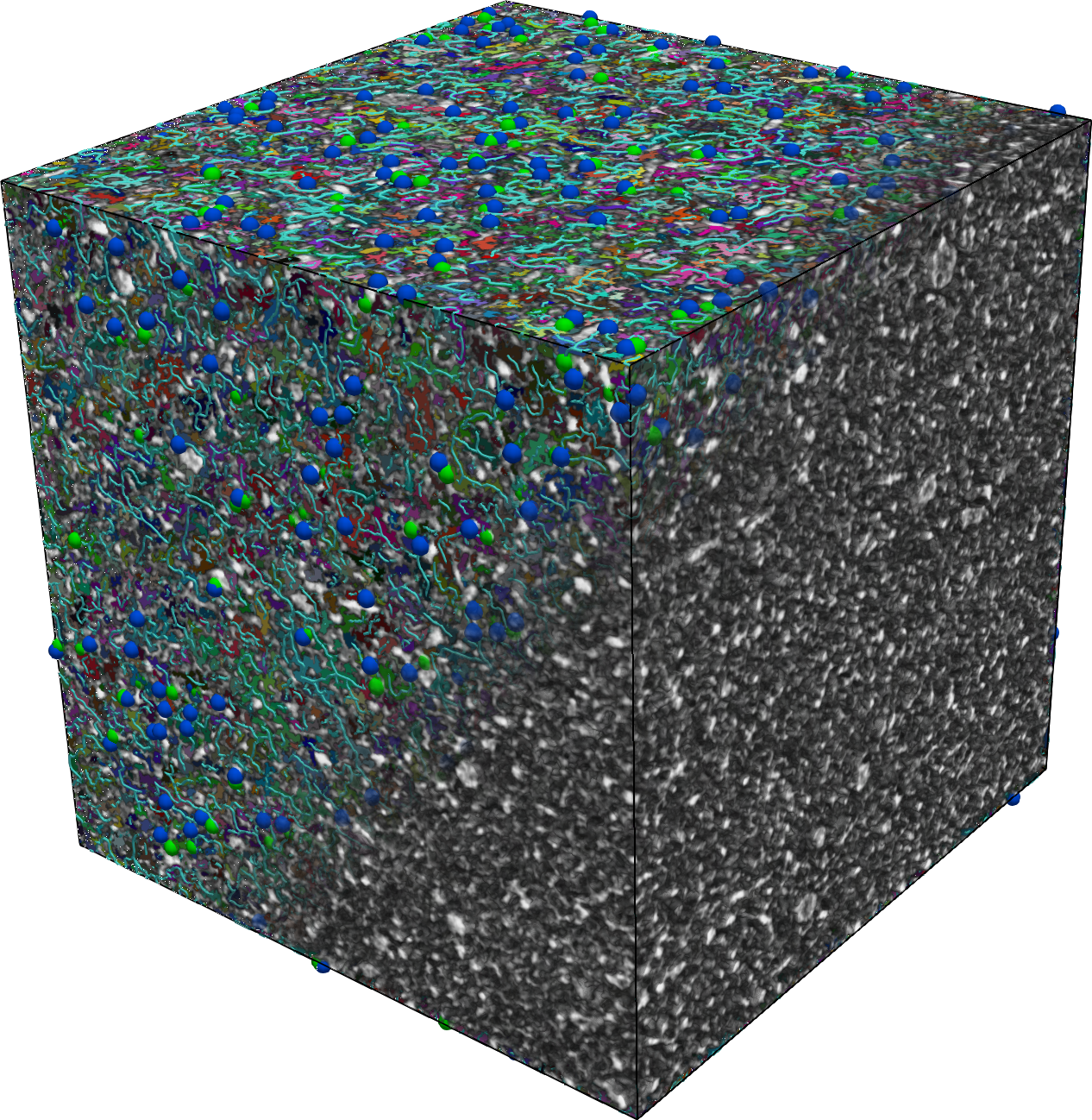}
        \caption { }
        \label{fig:HE_C}
    \end{subfigure}
    \caption{The materials used in our experiments. For each image, the top-left shows the decomposition into pores, the minima of the distance function corresponding to the pore basin (blue sphere), the 1-saddle on the interface between pores (green spheres) and the minimum-1-saddle-minimum paths (teal pipes) that define a throat between pores. The bottom-right of each image is a grayscale rendering of each image, with void space mostly transparent. The various sphere packings are shown on the top row: (a) Monodisperse-1, packed spheres with diameters from a bimodal distribution; (b) Polydisperse, with diameters from a lognormal distribution; and (c) Monodisperse-2, with uniform diameters. The materials on the bottom: (d) HE\_A, coarsely ground HMX crystals; (e) HE\_B, needle-like PETN crystals; and (f) HE\_C, finely-ground PETN crystals.}
\end{figure*}
\section{Results}
\label{sec_results}
 We first report the results of our PNM for three different sphere packing datasets with slightly different porosity levels, namely,  Monodisperse-1, Monodisperse-2 and Polydisperse as described in \autoref{sec:sphere_packing}. In \autoref{sec:he_materials}, we discuss in detail the results of our PNM on experimental micro-CT data for two different HEs, i.e., HMX and PETN, which are the materials of interest to our collaborators.

\begin{table*}
    \footnotesize
    \centering
    \begin{tabular}[t]{|c||c|c|c||c|c|c||c|c|c|c|c|c|}
        \multicolumn{1}{c}{\ } &  \multicolumn{3}{c||}{\textbf{Exp/Fisher measurement}} & \textbf{Isosurface} &  \multicolumn{2}{c||}{\textbf{Computed Conductance}} & \multicolumn{3}{|c|}{\textbf{ Fisher vs Computed}} & \multicolumn{2}{c|}{\textbf{Relative Order}} \\
        \hline
        Sample & $\epsilon$ & $S_f (m^2)$ & $\eta C_f (m^3)$ & $S (m^2)$ & $\eta C_{iso} (m^3)$ & $\eta C_{pnm} (m^3)$ &  $\frac{S_f}{S}$ & $\frac{C_f}{C_{iso}}$ & $\frac{C_{f}}{C_{pnm}}$ &  PNM & Fisher \\
        \hline
        \multicolumn{12}{c}{\  } \\
        \hline
        Monodisperse-1 & 0.6 & 6.95E-07 & 1.58E-17 & 7.06E-07 & 1.53E-17 & 0.63E-17 &  0.98 & 1.03 & 2.50 & 1 & 1 \\
        \hline
        Polydisperse & 0.53 & 6.09E-07 & 1.42E-17 & 5.80E-07 & 1.56E-17 & 0.53E-17 &  1.05 & 0.91 & 2.66 & 2 & 2 \\
        \hline
        Monodisperse-2 & 0.56 & 7.72E-07 & 1.00E-17 & 7.56E-07 & 1.04E-17 & 0.41E-17 & 1.02 & 0.96 & 2.43 & 3 & 3 \\
        \hline
         \multicolumn{12}{c}{\  } \\
       \hline
        HE\_A & 0.503 & 1.88E-04 & 0.58E-17 & 0.963E-04 & 2.21E-17 & 0.42E-17 &1.95 & 0.29 & 1.39 & 2 & 2 \\
        \hline
        HE\_B & 0.480 & 1.02E-04 & 1.71E-17 & 0.686E-04 & 3.78E-17 & 0.56E-17 & 1.49 & 0.45 & 3.02 & 1 & 1  \\
        \hline
        HE\_C & 0.483 & 2.03E-04 & 0.44E-17  & 0.104E-04 & 1.68E-17 & 0.24E-17 & 1.95 & 0.26 & 1.81 & 3 & 3 \\
        \hline
    \end{tabular}
    \caption{Comparison of Fisher measured flow-permeable surface area and conductance vis-a-vis MSC-PNM for three different sphere packing distributions. The left table contains the experimental measurements of the material: porosity $\epsilon$ is computed from measured density, mass, and volume of the packed sample; the surface area  $S_f (m^2)$ and the conductance $\eta C_{f} (m^3)$ are calculated from the volumetric flow rate measured by the Fisher apparatus, by~\autoref{eqn:carmankoseny}. The middle table lists the flow-permeable surface area computed using clipped isosurface $S (m^2)$, the conductance estimated from this surface area, $\eta C_{iso} (m^3)$, by applying~\ref{eqn:carmankoseny}, and the conductance computed with our PNM, $\eta C_{pnm} (m^3)$. The right table shows the factors between the experimentally measured and computed surface areas, $\frac{S_f}{S}$, the conductance derived from the surface area, $\frac{C_f}{C_{iso}}$, and the PNM conductance, $\frac{C_{f}}{C_{pnm}}$. %The isosurfaces computed for the sphere packings predict the measured surface area well, while the isosurface area for the HE materials is a consistent underestimate. 
    The relative order between materials, ordered by conductivity from most to least, matches between the spheres and between the HE materials.}
    \label{Tab:Spheres}
\end{table*}

\subsection{Sphere Packing} \label{sec:sphere_packing}
%The Carman–Kozeny model assumes that the solids are spheres with an average effective diameter $d_s$ \cite{Allen}. 
Sphere packing datasets have been commonly used in the literature as simple proxies for more complex morphologies and also because they lend themselves to simpler PNM construction and validation \cite{Hu2012}, \cite{Gostick2017}, \cite{Ulrike2014}. Furthermore, the Carman–Kozeny equation has been verified in this realm, relating surface area and flow properties to the effective diameter $d_s$ of spherical particles~\cite{Allen1997}. %In contrast to extracting a Voronoi graph of the distance transform of the spheres and comparing it to our pore network graph using graph matching, we chose to 
We evaluate the effectiveness of our PNM by comparison with experimentally measured surface area and volume flow rate for three different sphere packing distributions using the Fisher apparatus. 
%We evaluatse the effectiveness , by comparing our results with the Fisher measurements. 

The Monodisperse-1 and 2 datasets have all particles of either a single diameter or one of the two fixed diameters \autoref{fig:mono-1}, \autoref{fig:mono-2}, and the Polydisperse dataset has particles following a polydisperse distribution of diameters \autoref{fig:poly}. The micron-sized spherical particles with volume-weighted size distribution measured via laser diffraction were packed into a Fisher apparatus sample holder, and the flow-permeable surface area ($S_f$) and the volume flow rate was experimentally measured. We convert the volume flow rate into Fisher conductance ($C_f$) using \autoref{eqn:carmankoseny} for comparison. The experimental results are reported in \autoref{Tab:Spheres}(top). The samples were also imaged using micro-CT at $0.05\mu$m resolution, and the flow-permeable surface area and the volume flow rate for each of the three samples were computed using our PNM.

\autoref{Tab:Spheres}(top) tabulates the results of our PNM, as well as comparing the experimental and computed results. First, we notice that our PNM underestimates the flow-permeable surface area for all three datasets by a factor of 0.98-1.05. While the computed area is close, the variation seen is not unexpected: the stair-case digitizing of the smooth spherical boundary adds to the computed isosurface; the image resolution merging the boundaries of adjacent spheres subtracts from the computed surface area.

Our computed conductance $C_{pnm}$ underestimates the Fisher measured conductance, $C_f$ by a factor of 2.43-2.66 and $C_{iso}$, conductance computed using the isosurface area is approximately the same as the Fisher measured conductance. Although our model underestimates the conductance ($C_{pnm}$), the factors are close to one another for different porosity and particle size distributions, which is a promising outcome. In \autoref{sec:aspect_factor} we discuss in detail why this underestimate is consistent with expectations. We also note that our PNM accurately orders the materials in terms of conductivity (most to least) for the different sphere packings vis-a-vis the Fisher apparatus.

% \begin{figure}
%     \begin{subfigure}[h]{0.5\columnwidth}
%         \centering 
%         \includegraphics[trim=600 50 600 150,clip, width=\columnwidth]{figures/spheres_mono_seg.png}
%     \end{subfigure}
%     \begin{subfigure}[h]{0.5\columnwidth}
%         \centering 
%         \includegraphics[trim=600 50 600 150,clip, width=\columnwidth]{figures/spheres_mono_interface.png}
%     \end{subfigure}
% \end{figure}
% \begin{figure}
%     \begin{subfigure}[h]{0.5\columnwidth}
%         \centering 
%         \includegraphics[trim=600 50 600 150,clip, width=\columnwidth]{figures/spheres_lognorm_arcs.png}
%     \end{subfigure}
%     \begin{subfigure}[h]{0.5\columnwidth}
%         \centering 
%         \includegraphics[trim=600 50 600 150,clip, width=\columnwidth]{figures/spheres_lognorm_seg.png}
%     \end{subfigure}
% \end{figure}

\subsection{High-Explosive Materials} \label{sec:he_materials}
Material of two explosive molecules, HMX and PETN, were prepared, and the flow-permeable surface area and the volume flow rate was experimentally measured using the Fisher apparatus. HMX crystals tend to be hexagonal and relatively \textit{short}; thus they appear round in the HE\_A dataset \cite{miller2001review}, \cite{Cobbledick} \autoref{fig:HE_A}. On the other hand, PETN has been shown to have a higher aspect ratio (needle-like) crystals at elevated temperature during crystal growth and rounder crystals at lower growth temperatures \cite{ZEPEDARUIZ2006461}. These two distinct PETN grain shapes are represented in the experimental datasets HE\_B (needle-like) \autoref{fig:HE_B} and HE\_C (rounder) \autoref{fig:HE_C}. The three materials have similar porosity and were roughly packed to the sample density for the Fisher experiment. The experimental results are reported in \autoref{Tab:Spheres}(bottom). The samples were also imaged using micro-CT at $0.57\mu$m resolution, and the flow-permeable surface area and the volume flow rate for each of the three samples were computed using our PNM.

\autoref{Tab:Spheres}(bottom) tabulates the results of our PNM, as well as comparing the experimental and computed results. For the HE materials, our PNM underestimates the flow-permeable surface area by a factor of 1.49-1.95. The factor is higher than that for the sphere packing dataset, which is not unexpected; the roughness of the HE material, coupled with an order of magnitude lower resolution of the micro-CT, amplifies the smoothing of the solid/void interface. The percentage of total volume available for flow is 98\% for HE\_A, HE\_B and 100\% for HE\_C. This result was surprising for the domain experts, as the high-aspect shapes were expected to increase the likelihood of dead-end formation.

Our computed conductance $C_{pnm}$ underestimates the Fisher measured conductance, $C_f$ by a factor of 1.39-3.02, with HE\_B being the \textit{most} underestimated, despite being the highest-flow. The conductance computed using our flow-permeable isosurface, $C_{iso}$ is an overestimate by a factor of 0.26-0.45, which is not surprising given the underestimate of the surface area.  Similar to the sphere packing dataset, although our model underestimates the flow-permeable surface area ($S$) and the conductance ($C_{pnm}$), the factors are within the acceptable range for the different HE materials, which is an encouraging result. We also note that our PNM accurately orders the materials in terms of conductivity (most to least) vis-a-vis the Fisher apparatus.

\subsection{Empirical Bounds for Conductance}
% \item For simulated and imaged data, we use our approach to compute wetted surface area and conductance
Our conservative model for estimating the conductance detailed in the \autoref{sec_approach} means we expect to underestimate it computationally, which we also observed with our experiments. So the resistive network conductance can be used as a \textit{lower} bound on the material's conductance. Using this conductance in \autoref{eqn:carmankoseny} we can get an \textit{upper} bound of the flow-permeable surface area. Similarly, the resolution of the micro-CT w.r.t the grain/pore size of the materials means we expect to underreport the isosurface area (and therefore can be a \textit{lower} bound on flow-permeable isosurface area). Using our underreported isosurface area and \autoref{eqn:carmankoseny}, we, therefore, get an \textit{upper} bound on the material's conductance.

Our experimental results for the HE materials show that dead-ends are not a significant factor in the analysis of the HMX and PETN crystal materials. Note, a similar analysis could tell if the same holds true for: elastic grains, lower porosity, fractured materials. Our initial results from both the sphere packing and the high-explosives data build confidence that the predicted conductance from the PNM is related to the experimentally measured flow. With sufficient observed samples, it may be possible to build a predictive model using the PNM to avoid the Fisher experiments. Based on the promising preliminary findings using our approach, our material science collaborators are planning to produce more measurements varying porosity and flow rates in packed spheres and other particle shapes (e.g., non-uniform ‘grains’, ‘needle’ like crystallites, etc.), as well as produce an experimental measurement of additional aged and unaged HE samples that will be used to build a prediction model to predict material initiation performance.

\section{Discussion}
\label{sec_discussion}

We investigate the differences between experimentally measured results and our virtually measured ones. First, we discuss the factors we expect to impact the quality of the virtually measured flow-permeable surface area and computed volume flow rates and evaluate them with respect to the hyper-parameters of our PNM approach, such as image processing methodology, resolution, and persistence simplification threshold. Finally, we explore potential corrections using the computed tortuosity. 

\paragraph{Experimental error} Although we take the experimental measurements to be the ground truth for the imaged samples in our study, our collaborators noted that individual Fisher apparatus measurements could vary by up to 20\% for the same material -- due to differences in packing and sample preparation alone. This is especially true for the very small sample masses used in the micro-CT measurements, where full statistical averaging over flow paths may not be attained.

\subsection{Limitations of imaging}
It should be noted that the mean free path of air, 30-50 nanometers, is smaller by 1 order of magnitude than the micro-CT imaging resolution, at ~0.5 microns/voxel. Therefore, air can fit through cracks between grains that the image cannot even represent. An isosurface on such an image will necessarily under-estimate the total surface area of all grains. However, due to viscosity and zero fluid velocity at the solid/void interface, the pore structure that \textit{is} visible in the micro-CT is expected to dominate volumetric flow rates and flow-permeable surface area.

The grain roughness similarly is a dominant factor in flow calculation. For instance, simply smoothing the micro-CT image with a Gaussian kernel of radius$=1$ caused a nearly 20\% jump in the computed volume flow rates for HE\_A, HE\_B, and HE\_C. Although we used a simple threshold to determine solid/void in the imaged volumes, it is likely that noise in the micro-CT acquisition and reconstruction (due to beam hardening and other artifacts) causes artifacts that impact the flow rates. Furthermore, it is expected that the micro-CT resolution simultaneously hides very small features or roughness. We plan on re-applying our computational approach on micro-CT acquired at double the resolution.

\subsection{Doubling image resolution} \label{sec:topo_analysis}
In the original micro-CT images, we measured a median throat radius of just 1-2 voxel units ($vu$), which was consistent with the Fisher apparatus' measured value of 1.996$vu$. The 6-connectivity of the discrete gradient of a discrete sampling of a distance function means that a persistence simplification threshold of \textit{at least} l$vu$ needs to be applied to simply remove spurious critical points, and hence pore-throat-pore connections, introduced by the computation. Empirically, applying a 1$vu$ simplification threshold overly merged adjacent pores. Furthermore, as the conductance of a throat is based on discrete quadrilaterals, the small radius throats effected a poor sampling of the cross-section surface area. By doubling the mesh resolution in each dimension (using trilinear interpolation to re-sample images), the average radius doubles to 2-4$vu$ -- while the persistence threshold needed to remove artifacts remains 1$vu$. Furthermore, each throat is sampled with 4x more samples, leading to overall more accurate cross-section surface area and hence conductance. We evaluated a 2x and 4x magnification for each sample and the estimated measures of volume flow rate and total surface area were only marginally different for 2x and 4x, meaning 2x was sufficient.

The time to compute the discrete gradient for a $400^3$ image block resampled to $800^3$, on a 6 core, 3.5GHz Intel i7-5930K CPU, 64GB RAM, was 2.5 minutes for both spheres and HEs. Building the MS-complex, extracting throats and assigning conductances, and finally solving the resistor network took an additional 7 minutes for the spheres and 13 minutes for the HEs. The overall longer execution time for the HE materials is due to the increased topological complexity.

\begin{figure}
    \centering 
    \includegraphics[scale=0.11]{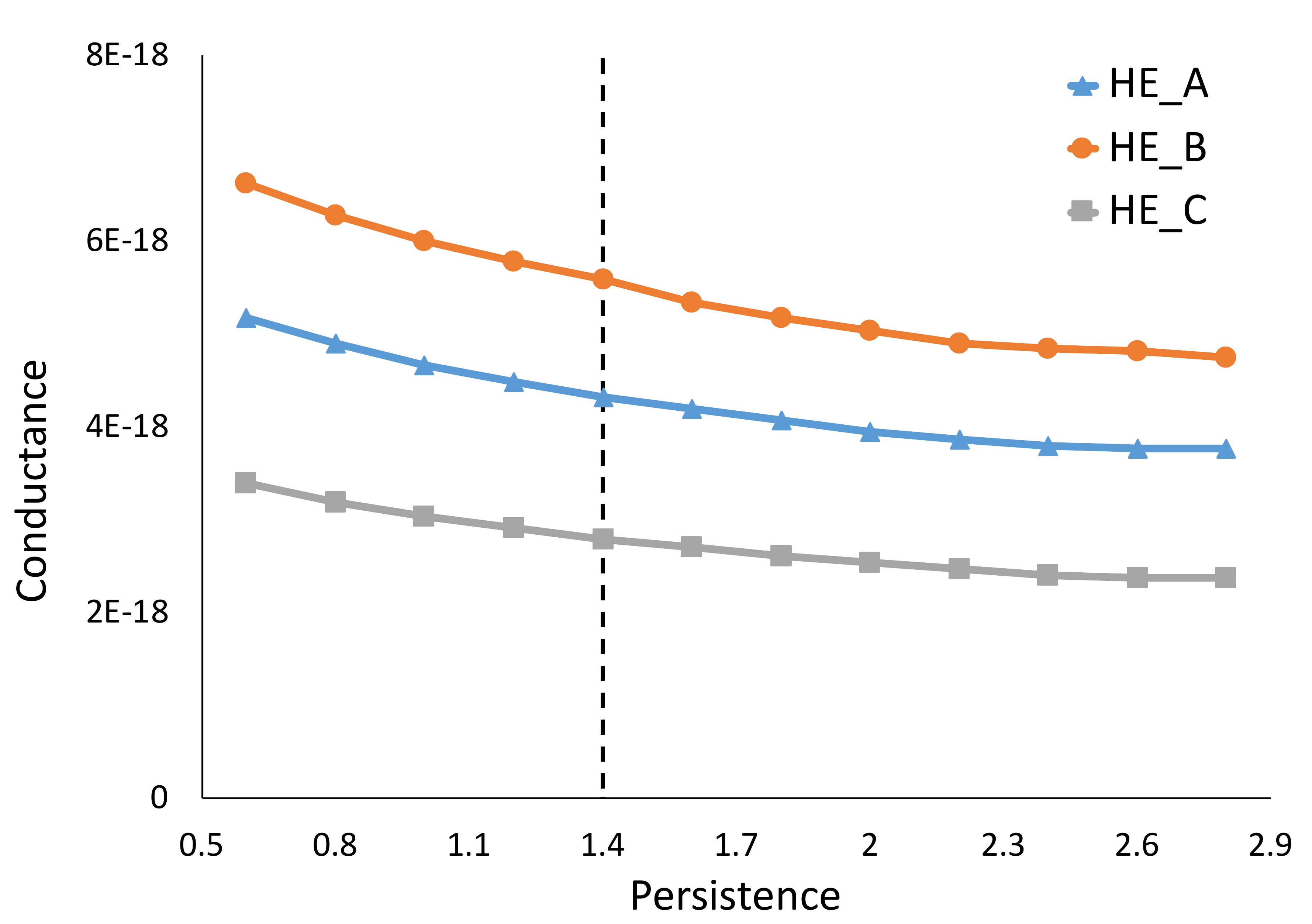}    
    \includegraphics[width=\columnwidth]{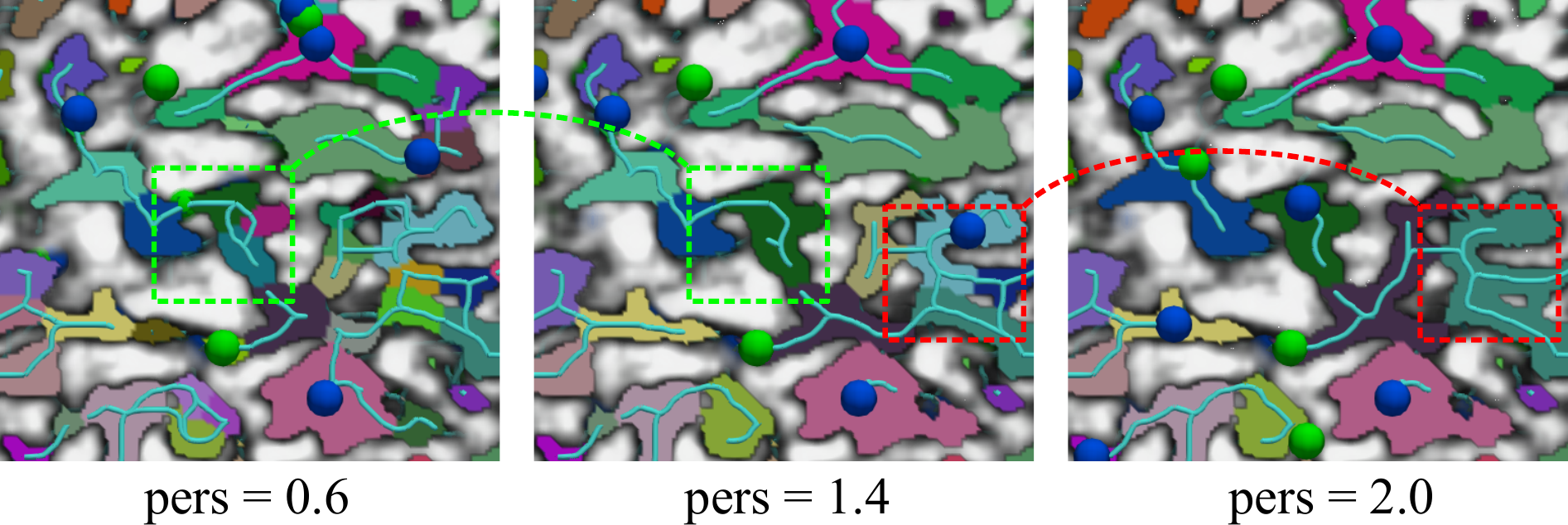}
    \caption{As the persistence simplification threshold is increased, neighboring pores are increasingly merged, while throat paths are extended. The bottom row shows three simplification levels. The green boxes highlight where increasing the threshold from 0.6 to 1.4 merges over-segmented regions. Further increasing the threshold to 2.0 begins to overly merge regions, highlighted by the red boxes. Here, the plot of computed effective conductance for HE\_A, HE\_B, and HE\_C show a similar response to increasing the persistence. We take the approach of setting the threshold to the minimum needed to reliably remove discretization artifacts at 1.4$vu$. The similarity of the curves indicates that the persistence threshold will likely not be a significant factor in the sensitivity of the outcome.}
    \label{fig:persistence_conductance}
\end{figure}

\subsection{Persistence threshold and stability} \label{sec:topo_pers}
Increasing the persistence simplification threshold merges pores and removes narrow throats from the PNM. We measured the effective conductance of each material as persistence was swept from 0.6 to 2.8, a range that encompasses known over-segmentation (due to discretization artifacts) and empirical under-segmentation, merging pores into tortuous shapes. ~\autoref{fig:persistence_conductance} shows that the conductance measured for all 3 materials respond similarly to changes in the persistence threshold -- indicating that persistence was not a significant factor in the sensitivity of the outcome. We set the persistence threshold to 1.4$vu$ for all 2x supersampled images, and provide the user with a visualization of the voids in a range bracketing this threshold, for visual verification. 

A significant simplifying assumption we have made in our approach is that the most restrictive throat between pores coincides with the ascending manifold of the 1-saddle separating the basins of the minima. 
While this empirically holds true for materials with convex grains (e.g. sphere packings), materials with varied grain morphology can give rise to longer, more tortuous connections, where slight perturbations in the image could ``move'' the 1-saddle, and thus, the representative throat interface, potentially changing the conductance value. Again, the flatness of the conductance vs. persistence curves in ~\autoref{fig:persistence_conductance} provides evidence that the calculation is stable with respect to perturbation of throat location. For instance, the PNM for  HE\_A has 9.3x the number of throats at persistence 0.6 as at persistence 2.8, while the conductance was only 1.38x higher, and the conductance reduces monotonically; this indicates that the computed conductance is not sensitive to the exact choices for throat interface, and on aggregate, the saddles retained \textit{do} correspond to more restrictive interfaces.

\subsection{Aspect factor} \label{sec:aspect_factor}
The Carman-Kozeny equation,~\autoref{eqn:carmankoseny}, has an experimentally, empirically determined factor $k$ that is called the aspect factor, measured to be $\sim$5 for most porous materials. This factor encompasses two terms, $k = k_0 k_1$, an aspect ratio and a tortuosity factor. The literature reports various tortuosity factors, usually defined as $k_1 = (L_e/L)^2$, where $L_e$ is the effective length a particle travels through the material, and $L$ is the length of the sample. The aspect ratio is usually reported as 2 for uniform sphere packings; however, this factor seems to be more measured rather than based on a model. Furthermore, the contributions of $k_1$ and $k_0$ are not well understood for non-spherical grains.  

The average flow particle path length, $L_e$, is given explicitly for our pore network model as the sum of the minimum-1-saddle-minimum path lengths weighted by the current on each path. \autoref{tab:tortuosity} reports these lengths normalized to the length of the side of the image cube. In the literature, $L_e$ is reported as closer to the shortest path length between the inflow and outflow. We compute the average shortest path lengths that connect a pore on the inflow to the outflow, denoted $L_{sp}$. We observe that our computed particle path length $L_e$ is a significant factor larger than $L_{sp}$. This is consistent with the expectation that, by defining the pipe length to be the complete path length between minima, we overestimate, during construction, the pipe length in the conductance calculation. Given the ambiguity regarding $k$, $k_1$, we explore using our tortuosity overestimation directly as a scaling factor to correct the overestimation. \autoref{tab:tortuosity} shows that multiplying the conductance for each material by $L_{corr}=L_e/L_{sp}$ reduces the underestimation of the conductance by our PNM. % \todo{add anything here?}

\begin{table}
    \footnotesize
    \begin{tabular}[t]{|c|c|c|c|c|c|}
            \hline
         \ &  \multicolumn{3}{c|}{Total, Ave. shortest path} &  \multicolumn{2}{c|}{Corrected Conductance}  \\
            \hline
         Sample & $\frac{L_e}{L}$ & $\frac{L_{sp}}{L}$ & $L_{corr}$  & $\eta L_{corr} C_{pnm}$ & $ \frac{C_f}{L_{corr}C_{pnm}}$ \\
        \hline
        \hline
        Monodisperse-1 & 1.74 & 1.23 & 1.41 & 0.89E-17 & 1.58\\
        \hline
        Polydisperse & 1.71 & 1.20 & 1.42 & 0.75E-17 & 1.87\\
        \hline
        Monodisperse-2 & 1.77 & 1.23 & 1.44 & 0.59E-17 & 1.69\\
        \hline
        \hline
        HE\_A & 2.55 & 1.44 & 1.77 & 0.74E-17 & 0.78\\
        \hline
        HE\_B & 2.56 & 1.46 & 1.76 & 0.99E-17 & 1.72\\
        \hline
        HE\_C & 2.72 & 1.51 & 1.80 & 0.44E-17 & 1.01\\
        \hline
    \end{tabular}
    \caption{For the packed spheres (top) and HE materials (bottom), we compute the average length traveled by an electron, $L_e$, the average shortest path between inflow and outflow $L_{sp}$, both normalized by the length of the image cube. The factor between these two $L_{corr}$ is used to correct the conductance value and recompute the scale factor between measured and computed conductance.}
    \label{tab:tortuosity}
\end{table}

\begin{figure}
    \centering 
    \includegraphics[width=0.95\columnwidth]{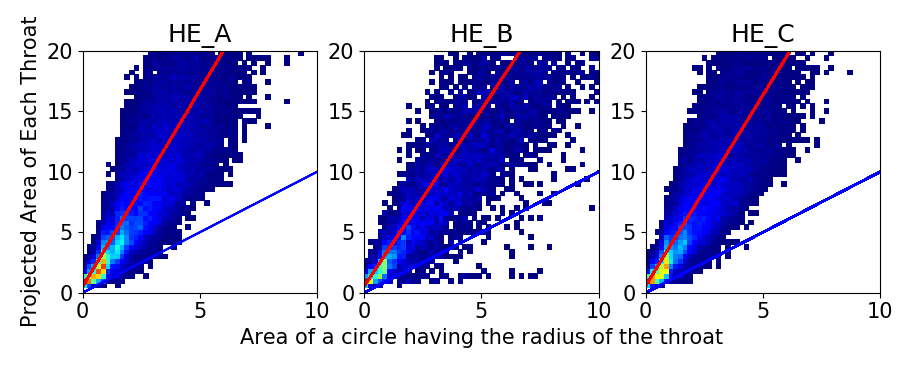}
    \includegraphics[scale=0.25]{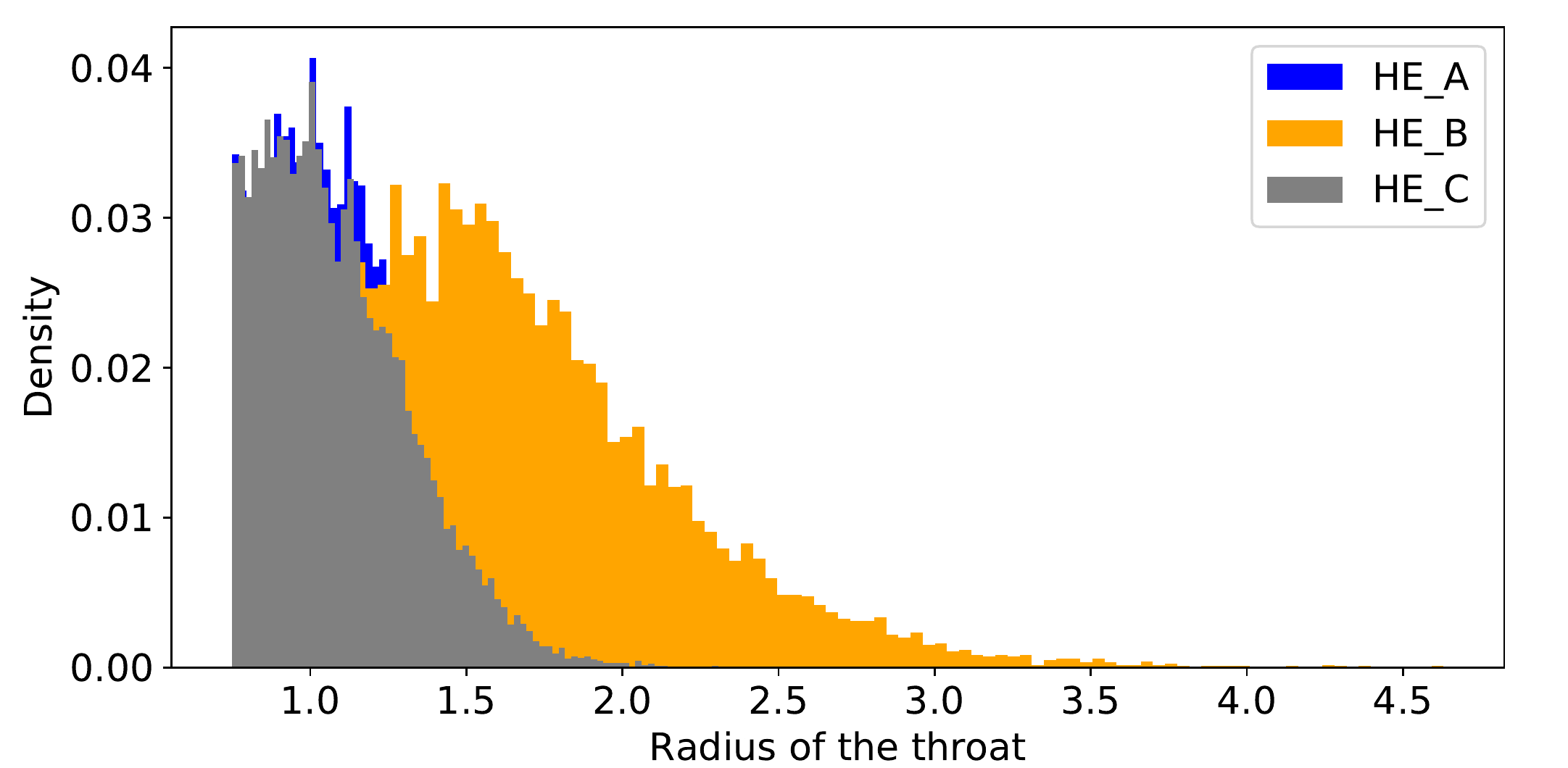}
    \caption{(Top) A 2D histogram of the projected area of each throat to the area of a circle having the radius of the throat. The (blue) line $ y = x$ would be expected to fit the plot if each throat were circular in cross-section. Despite differences in grain shapes, a linear fit yields to $y = $$\sim3x$ for each material. (Bottom) Histograms for each HE material of the maximum absolute radius of a throat. The histograms for the 2 materials of HMX (HE\_A, and HE\_C) are nearly identical, despite the different coarseness of grinding in preparation of the samples.  }
    \label{fig:ratio}
\end{figure}

Finally, we test the hypothesis that the throats in the different HE materials vary in the aspect ratio of their cross-sections. In \autoref{fig:ratio}, we plot a histogram of the area a perfectly circular cross-section would have given using the measured radius from the distance field to the measured projected area. Perfectly circular cross-sections would appear along the line $y=x$. More elongated shapes have higher aspect ratios. Surprisingly, the aspect ratios for throats for all three materials followed the same fit; the throats are similarly shaped despite the different grain shapes. We also plot the histogram of throat radii, normalized by the throat count, and observe that HE\_B has larger radius throats, as expected due to the larger particle sizes.

\section{Conclusions and Future Work}
\label{sec_conclusions}

The new approach in constructing a PNM based on topological techniques has demonstrated that it is feasible to construct a virtual Fisher apparatus which, if adequately calibrated, would represent a significant breakthrough in the study of porous structures. 
Nevertheless, while self-consistent and in line with the expectations of our subject matter experts, the quantitative results are noticeably different from the current experimental data. There are two likely causes for this discrepancy that we will explore in future work. 
First, the resolution of the current images is too coarse to allow a fully accurate segmentation of grains and, more importantly, too coarse to estimate the surface roughness of the grains. Consequently, we are prone to underestimating the surface area when using the isosurface-based estimation.However, these issues can be addressed with higher resolution CT scans, which, based on our results, are currently being planned. 
The field of view of these scans will be necessarily smaller, but the expectation is that changes in surface roughness and tortuosity can be estimated from smaller samples and transferred to large systems. 
Second, our simplified PNM assumes a quadratic velocity profile and selects a single representative throat surface for computing the conductivity of a pipe, thereby grossly underestimating it. We expect to improve the accuracy of the PNM prediction by adapting this conductance based on the local radius of the pipe and devising methods to better estimate pipe length.
A more interesting potential source of error is the approximation error inherent in the Carman-Kozeny equation. 
It is based on empirical observations and simple systems amenable to analytic solutions and is expected to be incorrect for the decidedly non-spherical grains of HE materials. 
An adequate correction for the computed surface area and the true tortuosity of these systems would provide crucial insights into the underlying physics and could lead to either updated parameters for the Carman-Kozeny equation or even an entirely new formulation. 
Ultimately, a major goal of this work is to provide descriptive numbers that can be used to build a material performance model; in this context, we also plan to investigate the relationship with percolation thresholds~\cite{broadbent_hammersley_1957}.

\acknowledgments{This research is supported in part by the Department of Energy under Award Number(s) DE-FE0031880 and the Exascale Computing Project (17-SC-20-SC), a collaborative effort of the U.S. Department of Energy Office of Science and the National Nuclear Security Administration, and in part by NSF OAC award 1842042, NSF OAC award 1941085, and NSF CMMI award 1629660. The work at LLNL was performed under the auspices of the U.S. Department of Energy by Lawrence Livermore National Laboratory under Contract DE-AC52-07NA27344.}

\bibliographystyle{abbrv-doi}

\bibliography{ms, will-auto-export, msc-vrnt}
\end{document}